\documentclass{emulateapj}
\usepackage{apjfonts}
\usepackage{natbib}
\usepackage{graphicx}

\newcommand{\etal}{{\rm et al.}}
\newcommand{\eg}{{\rm e.g.}}

\newcommand{\ie}{{\rm i.e.}}
\newcommand{\etc}{{\rm etc.}}
\newcommand{\kms}{km s$^{-1}$}

\newcommand{\msol}{\ensuremath{\ \mathrm{M_{\odot}}}}

\newcommand{\feh}{\ensuremath{[\mathrm{Fe/H}]}}
\newcommand{\afe}{\ensuremath{[\alpha\mathrm{/Fe}]}}

\newcommand{\s}[2]{$\mathcal{S}_{#1,#2}$}
\newcommand{\jz}{\ensuremath{j_z}}
\newcommand{\jcirc}{\ensuremath{j_\mathrm{circ}}}
\newcommand{\zmed}{\ensuremath{z_\mathrm{med}}}
\newcommand{\cir}{\ensuremath{\epsilon}}
\newcommand{\jcrit}{\ensuremath{\epsilon_\mathrm{crit}}}
\newcommand{\Etot}{\ensuremath{E_\mathrm{tot}}}
\newcommand{\tform}{\ensuremath{t_\mathrm{form}}}
\newcommand{\rgc}{\ensuremath{r_\mathrm{gc}}}
\newcommand{\gcr}{\ensuremath{R_\mathrm{g}}}
\newcommand{\dgcr}{\ensuremath{\Delta R_\mathrm{g}}}
\newcommand{\rform}{\ensuremath{r_\mathrm{form}}}

\newcommand{\vzdisp}{\ensuremath{\sigma_{\mathrm{v_z}}}}
\newcommand{\vrdisp}{\ensuremath{\sigma_{\mathrm{v_r}}}}
\newcommand{\vznon}{\ensuremath{\sigma_{\mathrm{v}_{z,non}}}}
\newcommand{\vzom}{\ensuremath{\sigma_{\mathrm{v}_{z,om}}}}

\def\kmsmpc{\,{\rm km\,s^{-1}\,Mpc^{-1}}}
\def\Mo{{\rm M_\odot}}

\shorttitle{Disk Galaxy Assembly}
\shortauthors{Bird \etal}

\begin{document}

\title{Inside Out and Upside Down: Tracing the Assembly of a Simulated Disk Galaxy Using Mono-Age Stellar Populations}

\author{Jonathan C. Bird\altaffilmark{1,2,3,8}, Stelios
Kazantzidis\altaffilmark{1,2}, David H. Weinberg\altaffilmark{1,2}, \\ 
Javiera Guedes\altaffilmark{4}, Simone Callegari\altaffilmark{5}, Lucio
Mayer\altaffilmark{6}, \& Piero Madau\altaffilmark{7}}

\altaffiltext{1}{Department of Astronomy, The Ohio State University, 140 West 18th Avenue, Columbus, OH 43210}
\altaffiltext{2}{Center for Cosmology and Astro-Particle Physics, The Ohio State University, 191 West Woodruff Avenue, Columbus, OH 43210}
\altaffiltext{3}{Department of Physics and Astronomy, Vanderbilt University, 6301 Stevenson Center, Nashville, TN, 37235}
\altaffiltext{4}{Institute for Astronomy, ETH Z\"urich, Wolgang-Pauli-Strasse 27, 8093 Z\"urich, Switzerland}
\altaffiltext{5}{Anthropology Institute and Museum, University of Z\"urich, Winterthurerstrasse 190, CH-8057 Z\"urich, Switzerland}
\altaffiltext{6}{Institute for Theoretical Physics, University of Z\"urich, Winterthurerstrasse 190, CH-8057 Z\"urich, Switzerland}
\altaffiltext{7}{Department of Astronomy \& Astrophysics, University of California, 1156 High Street, Santa Cruz, CA 95064}
\altaffiltext{8}{VIDA Postdoctoral Fellow}

\begin{abstract} 
We analyze the present-day structure and assembly history of a high resolution
hydrodynamic simulation of the formation of a Milky Way (MW)-like disk galaxy,
from the ``Eris'' simulation suite, dissecting it into cohorts of stars formed
at different epochs of cosmic history.  At $z=0$, stars with $\tform < 2$ Gyr
mainly occupy the stellar spheroid, with the oldest (earliest forming) stars
having more centrally concentrated profiles.  The younger age cohorts populate
disks of progressively longer radial scale length and shorter vertical scale
height.  At a given radius, the vertical density profiles and velocity 
dispersions of stars vary smoothly as a function of age, and the superposition
of old, vertically-extended and young, vertically-compact cohorts gives rise to
a double-exponential profile like that observed in the MW. Turning to formation
history, we find that the trends of spatial structure and kinematics with 
stellar age are largely imprinted at birth, or immediately thereafter.  Stars 
that form during the active merger phase at $z>3$ are quickly scattered into 
rounded, kinematically hot configurations. The oldest disk cohorts form in 
structures that are radially compact and relatively thick, while subsequent 
cohorts form in progressively larger, thinner, colder configurations from gas 
with increasing levels of rotational support. The disk thus forms 
``inside-out'' in a radial sense and ``upside-down'' in a vertical sense. 
Secular heating and radial migration influence the final state of each
age cohort, but the changes they produce are small compared to the trends
established at formation. The predicted correlations of stellar age with 
spatial and kinematic structure are in good qualitative agreement with the 
correlations observed for mono-abundance stellar populations in the MW.
\end{abstract}

\section{Introduction}
\label{sec:intro}

Eggen, Lynden-Bell, and Sandage (\citeyear{Eggen62}; hereafter, ELS)
presented one of the earliest theoretical accounts of Milky Way
(MW) formation: 
stars formed during the gravitational contraction of a 
rotating gas cloud, with successive generations forming
successively more flattened, rotationally supported populations.
They argued that this picture could explain the observed correlation
between orbital eccentricity and chemical composition. Subsequent analytic models have
emphasized the dark matter (DM) halo's role as primary potential well and
baryon angular momentum as a governor of disk scale length, with SN-driven
outflows having a major impact on baryonic mass \citep[\eg,][]{White_Rees78,
Fall80,White91,Kauffmann93,Cole94,Mo98}. 
In the cold dark matter (CDM) scenario, hierarchical
structure formation leads to complex assembly histories whose effects can only
be fully assessed with numerical simulations, which have grown steadily in
computational and physical sophistication \citep[\eg,][to give just three
snapshots of a voluminous field]{Katz92,Navarro97,Brook12}. In this paper, we
analyze one of the highest resolution simulations ever run of a MW-like galaxy
formed from cosmological initial conditions to investigate its detailed
assembly history.

The physics governing the construction of each major galactic component is
likely to be different. Mergers and cannibalism of dwarf satellites may play a
major role in the formation of the stellar halo, perhaps accounting for most of
its mass \citep{Searle78,Bullock01,Bullock05}. The bulge could be the remnant
of early mergers \citep[\eg,][]{Hopkins10} or the result of internal processes
such as bar and bending instabilities that heat the inner disk over the life of
the galaxy \citep[\eg,][]{Debattista04,Debattista05}. Thin disk formation is
likely driven by gas dissipation and angular momentum conservation as in the
classic picture. However, bimodal distributions in vertical star counts
\citep[\eg,][]{Gilmore83}, kinematics \citep[\eg,][]{Bensby03}, and chemistry
\citep[\eg,][]{Lee11} in the solar neighborhood suggest a distinct thick disk.
Many theories of its origin have been proposed. The thick disk may arise from
the accretion of stars stripped from satellites \citep{Abadi03}, from a
satellite dynamically heating a pre-existing stellar disk
\citep[\eg,][]{Kazantzidis08, Villalobos08, Kazantzidis09}, from a gas rich
merger at early times \citep{Brook04}, from a clumpy and turbulent ISM at high
redshift \citep[\eg][]{Bournaud09}, or from stars migrating outwards from the
hot, inner disk \citep{Schonrich09, Loebman11}. 

Classic chemical evolution models describe disk galaxy formation using an
inside-out star formation and chemical enrichment history
\citep[\eg,][]{Matteucci89}. This approach is sensible; the inner disk is
thought to assemble first and form stars faster because of the high density of
accreted gas in the center of the galaxy's potential well. However, both
theoretical and observational studies over the last decade have emphasized the
potential role of radial migration as a mechanism to redistribute stars and
diversify the distribution of stellar chemical composition as a function of
radius \citep[\eg,][]{Sellwood02, Haywood08,
Roskar08b,Schonrich09,Casagrande11, Minchev12b}.  Radial migration
alleviates tension between models and observations of the local stellar
metallicity distribution and super metal rich stars in the solar annulus
\citep{Schonrich09}; it also naturally reproduces both the vertical density profile of stars and the chemical bimodality of the thin and thick disk \citep{Schonrich09,Schonrich09b}. Still, the contribution of radial migration to the dynamical evolution of
the disk, specifically thick disk dynamics, is currently debated in the
literature \citep[\eg,][]{Schonrich09b, Loebman11, Minchev12a, Roskar12b}. 

In this paper, we dissect a galaxy simulated in a fully cosmological context
into individual age cohorts of stars formed at different epochs. We study the
present-day spatial structure and kinematics of these cohorts, as well as their
assembly history and radial migration patterns. Our results are less directly
comparable to observations than chemistry-based investigations, but the age
cohorts are physically simpler than chemically-selected tracer populations and
avoid the uncertainties associated with numerical chemical enrichment
implementation \citep[\eg,][]{Shen10}. In this paper, age cohorts reveal the
expected inside-out growth of the disk. The simulated galaxy's formation can
also be described as ``upside-down'' in the sense that old stars form in a
relatively thick component, or are kinematically heated very quickly after
their birth.  Younger populations form in successively thinner disks. Stars
radially migrate throughout the galaxy's assembly, but migrators do not disrupt
the general kinematics-age trends put into place by the ``upside-down''
construction. There is some overlap of our results with those of
\citet{Brook12}, who study the buildup of a chemically-defined thin disk, thick
disk, and halo in the solar annulus of a less massive simulated disk galaxy; we
discuss a comparison with their work in Section~\ref{sec:discussion}.

Motivation for our study comes partly from analysis of the SDSS Sloan Extension
for Galactic Understanding and Exploration (SEGUE) \citep{Yanny09} G dwarf
sample by \citet{Bovy12b,Bovy12a,Bovy12c}. They dissect the sample into groups
spanning narrow ranges of \feh\ and \afe\ and discover that the physical
configuration of each mono-abundance population is  well described by a single
exponential function in both the radial and vertical directions; specifically,
more alpha-rich, iron-poor (old) populations have shorter scale lengths but
larger scale heights than alpha-poor, iron-rich (young) populations (see their
Figure 5). \citet{Bovy12b} use this result to contend that the MW's thick disk
is not a distinct structural component, a possibility raised by the mixture models of \citet{Nemec91,Nemec93}. The mono-abundance populations also
have simple, isothermal kinematics in the vertical direction \citep{Bovy12c}.
The assembly histories of the age-cohorts (Section~\ref{sec:evolution})
illustrate the simulation's fully self-consistent galaxy formation scenario
that culminates in galactic structure (Section~\ref{sec:current}) qualitatively
similar to the mono-abundance description of the MW.

\section{The Simulation} 

\label{sec:sim} The cosmological simulation employed
in the present study is part of the ``Eris'' campaign \citep{Guedes11} 
of hydrodynamical simulations of the formation of Milky Way
(MW)-sized disk galaxies. It was performed with the parallel, TreeSPH $N$-body
code \textsc{gasoline} \citep{Wadsley_etal04} in a {\it Wilkinson Microwave
Anisotropy Probe} 3-year cosmology \citep[$\Omega_M=0.24$, $\Omega_\Lambda=0.76$,
  $\Omega_b=0.042$, $H_0=73\,\kmsmpc$, $n=0.96$, $\sigma_8=0.76$; ][]{Spergel07}.

\begin{figure*}[!ht] 
  {\centering \includegraphics[width=\textwidth]{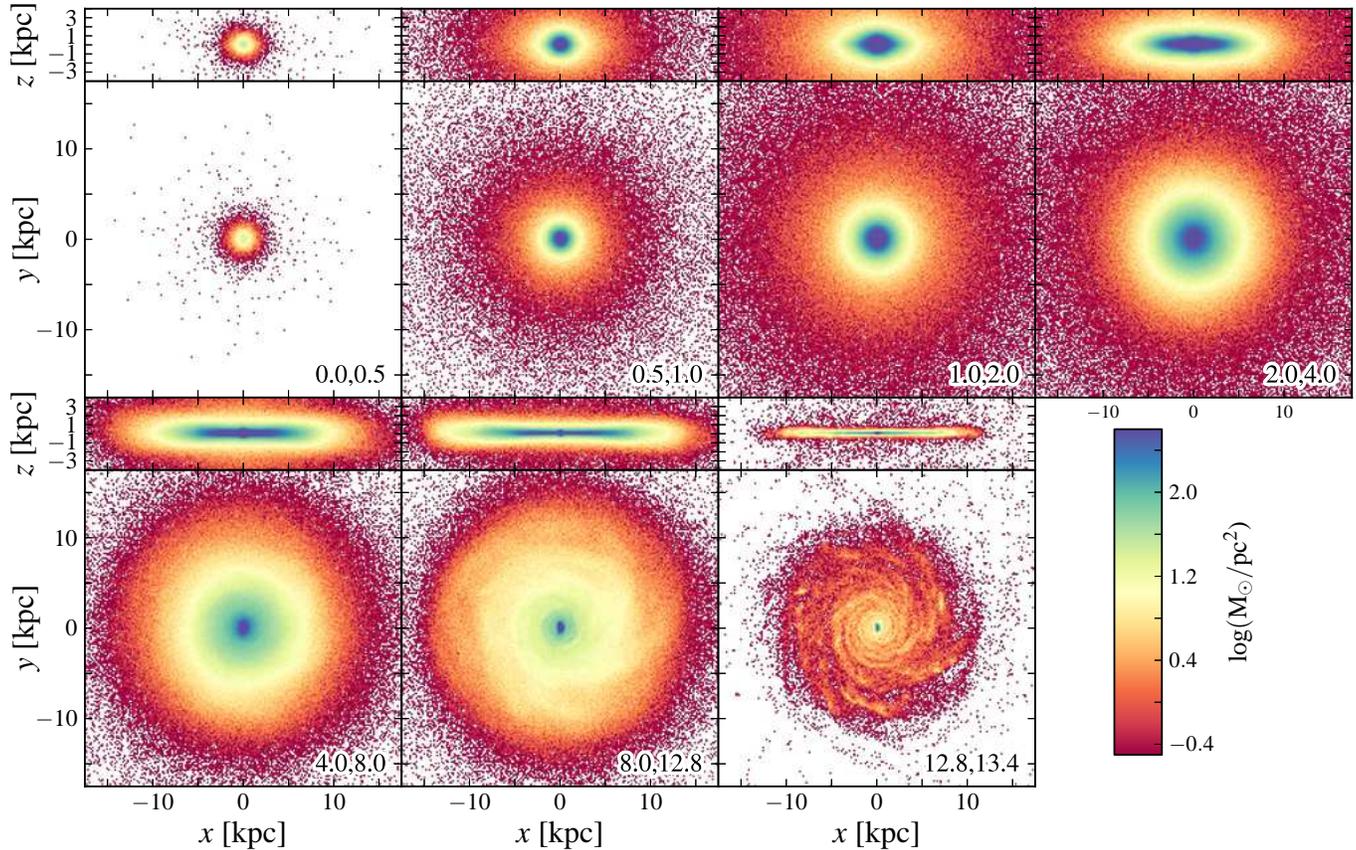}}
  \caption{\label{fig:400_den}The present-day surface density of stars in the
  simulated galaxy as a function of position and age. Each panel shows the face-on and
  edge-on views of a single age cohort (formation time bins, in Gyr after the
  big bang, are denoted in the bottom right corner of each panel); pixel color
  represents the logarithmic surface density at the pixel position (see
  colorbar). The coordinate system is such that the disk lies in the $xy$ plane
  and the galaxy's angular momentum vector is aligned with the $z$ axis. Age
cohort spatial structure smoothly evolves: old populations  are radially short
and vertically extended; younger cohorts with later formation times inhabit
progressively longer and thinner disks. }
\end{figure*}

All simulation parameters are identical to those of \citet{Guedes11},
except for the star formation efficiency parameter, which was equal to
$\epsilon_{\rm SF} = 0.1$ in the latter study, but it is $\epsilon_{\rm SF} =
0.05$  in the simulation employed here (see below). Specifically, we started
from a low-resolution ($300^3$ particles), DM-only simulation corresponding to
a periodic box of $90$~Mpc on a side. The target halo was identified at $z=0$
and was chosen to have a mass similar to that of the MW and a rather quiet late
merging history, namely to have experienced no major mergers (defined as
encounters with mass ratios $\ge 1/8$) after $z=3$. Three halos with masses
between $6\times 10^{11}$ and $8\times 10^{11}\,\Mo $ satisfied these
conditions in the (90~Mpc)$^3$ box, and we chose to re-simulate the one which
appeared to be most ``regular'' and isolated, with a virial mass of $M_{\rm
vir}=7.9\times 10^{11}\,\Mo$ at $z=0$.

New initial conditions were then generated with improved mass resolution (by a
factor $20^3$) on a nested grid in a Lagrangian sub-region of size $1$~Mpc,
centered around the chosen halo, using the standard ``zoom-in'' technique to
add small-scale perturbations. High-resolution particles were further split
into 13 million dark matter particles and an equal number of gas particles, for
a final dark and gas particle mass of $m_{\rm DM}=9.8\times 10^4\,\Mo$ and
$m_{\rm SPH}=2\times 10^4\,\Mo$, respectively.  The gravitational softening
length was fixed to $120$~pc for all particle species from $z=9$ to $z=0$, and
evolved as $1/(1+z)$ from $z=9$ to the starting redshift of $z=90$. At the
present epoch, the simulation contains $\sim 7$, $\sim 3$, and $\sim 8.6$
million dark matter, gas, and star particles, respectively, for a total of $N
\sim 18.6$ million particles within the virial radius ($R_{\rm vir} \sim
239$~kpc).

The version of \textsc{gasoline} used in this study includes Compton
cooling, atomic cooling, and metallicity-dependent radiative cooling
at low temperatures \citep{Mashchenko_etal06}. A uniform UV background
modifies the ionization and excitation state of the gas and is
implemented using a modified version of the \citet{Haardt_Madau96} spectrum.

Our star formation recipe follows that of \citet{Stinson_etal06},
which is based on that of \citet{Katz92}. Star formation
occurs when cold ($T<T_{\rm max}$), virialized gas that is part of a
converging flow reaches a threshold density ($n_{\rm SF}$). Gas
particles spawn stellar particles with a given efficiency
$\epsilon_{\rm SF}$ at a rate proportional to the local dynamical
time, $t_{\rm dyn}$,
\begin{equation}
d\rho_*/dt=\epsilon_{\rm SF} \rho_{\rm gas}/t_{\rm dyn} \propto \rho_{\rm gas}^{1.5}
\end{equation}
(i.e., locally enforcing a Schmidt law), where $\rho_*$ and $\rho_{\rm
gas}$ are the stellar and gas densities, respectively. In our
simulation, $T_{\rm max}=3\times 10^4$~K, $n_{\rm SF}=5$ atoms
cm$^{-3}$, and $\epsilon_{\rm SF}=0.05$.

Star particles are created stochastically with an initial mass $m_*
\sim 6.2\times 10^3\,\Mo$, and the gas particle that spawns the
new star has its own mass reduced accordingly. Gas particles are
deleted from the simulation once their masses fall below $\sim 4\times
10^3\,\Mo$ (with their masses and metals being distributed in
neighboring gas particles). A newly formed star particle represents a
simple stellar population with its own age and metallicity and a 
\citet{Kroupa93} initial stellar mass function.

Feedback from supernovae (SNe) is treated using the blast-wave model
described by \citet{Stinson_etal06} , which is based on the analytic
treatment of blastwaves described by \citet{McKee_Ostriker77}.  Each
Type-II SN deposits metals and a net thermal energy of $\epsilon_{\rm
SN} \times 10^{51}\,$ergs into the nearest neighboring gas particles,
and the fraction of energy that couples to the interstellar medium was
set to $\epsilon_{\rm SN}=0.8$ (the same value was adopted in previous
cosmological simulations; see, for example,
\citealt{Governato_etal07}). The heated gas has its cooling shut off
for a timescale set by the local gas density and temperature
and by the total amount of energy injected \citep{Stinson_etal06}. The
energy deposited by many SNe adds up to create larger, hot bubbles and
longer cooling shutoff times. No feedback from an
active galactic nucleus was included in the simulation.

Our high mass and force resolution enables us to adopt a density threshold for
star formation that is $50$ times higher compared to those employed in previous
lower-resolution studies. The local Jeans length corresponding to our density
threshold and $T = 10^3 K$ is resolved with more than $5$ SPH smoothing
lengths, and thus artificial fragmentation is prevented \citep{Bate_Burkert97};
we note that very few gas particles inside the virial radius ever cool below
$10^3 K$.  Similarly, the Jeans mass at our threshold density is resolved with
at least $100$ $m_{\rm SPH}$, where $m_{\rm SPH}$ denotes the mass of the SPH
gas particle. Lastly, we stress that the star formation density threshold
adopted here is high enough to allow the development of a clumpy, inhomogeneous
ISM, which serves as a vital element for the formation of realistic disk
galaxies \citep{Guedes11}.

Our simulation produces a galaxy that at $z=0$ has structural properties nearly
identical to those of the base "Eris" run described by \citet{Guedes11}
(for example the bulge-to-disk ratio and the radial scale length of the disk
differ by less than $10\%$ in the two simulations; see \citealt{Mayer12}).
Although we defer a detailed comparison of the two simulations  to future work,
we stress that the rotation curve of our resulting galaxy is even flatter than
that of \citet{Guedes11}, perhaps because the nuclear bar is
weaker and thus less dense, providing an even better match to the rotation
curve of a typical late-type Sb/Sbc spiral galaxy.

\begin{figure}
 \includegraphics[width=3.5in]{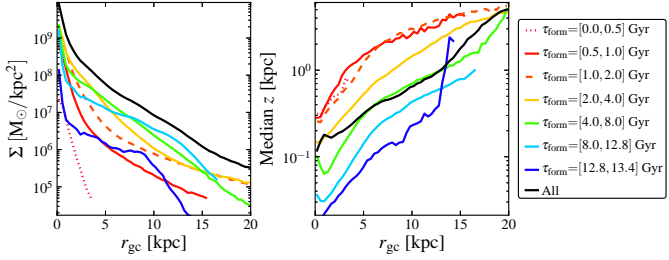}
 \caption{\label{fig:400_pos} The radial profiles of surface mass density (left
 panel) and median height above the disk plane (right panel) for all age
 cohorts at redshift zero. Note \rgc, used throughout this paper, refers to
 galactocentric radius. The color of each radial profile indicates cohort
 formation time; colors progress from red for old stars to blue for young stars
 (see legend for details; in addition to color, line types help differentiate
 the oldest cohorts). We also plot the surface mass density and median height of all stars (thick, black lines). All radial profiles in this paper have $0.37$ kpc
 bins; bins with $n<10$ particles are left blank. Younger cohorts populate more
 disk-like configurations and reside closer to the plane than their older
 counterparts. }
\end{figure}

\section{The Present-Day Galaxy} \label{sec:current}
\begin{figure*}[ht] 
\includegraphics[width=\textwidth]{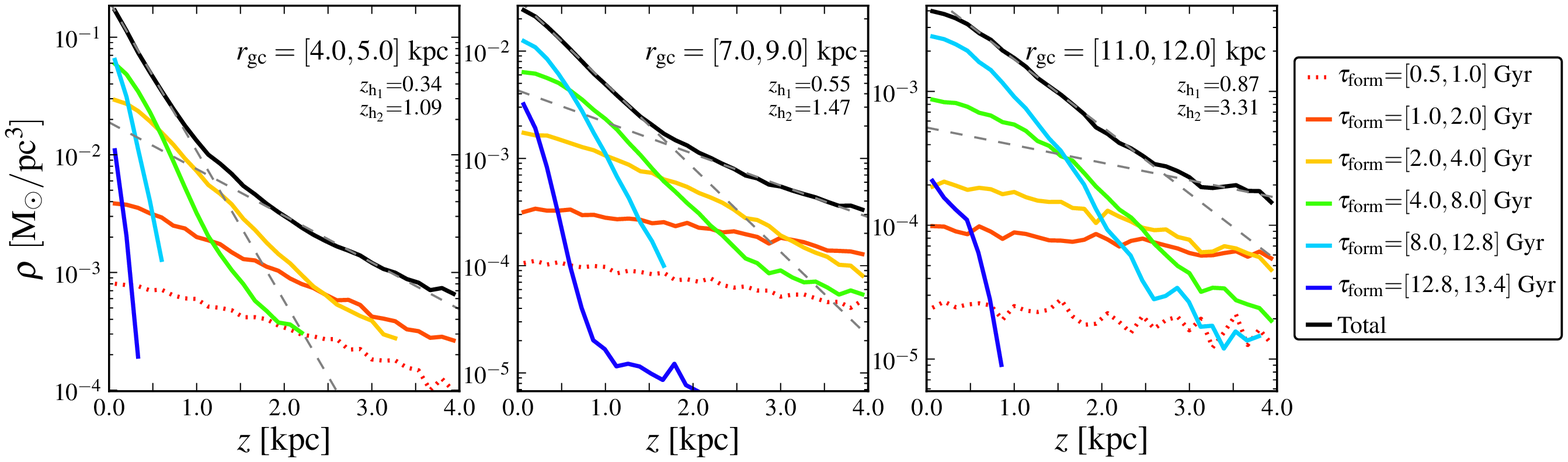} 
\caption{\label{fig:400_vp} Vertical mass density profiles of the age cohorts
at the three labeled radial annuli in the disk (from left to right: $\rgc=[4.0,5.0]$,
$[7.0,9.0]$, and $[11.0,12.0]$ kpc). The $y$ axis for all panels is the mass
density in \msol\ pc$^{-3}$. The $x$ axis shows the same range in height above
the plane (in kpc) for all panels. The color-coding is the same as in
Figure~\ref{fig:400_pos} (see legend). Thick, black lines represent the total
vertical mass density profile in each annulus. We fit two exponential profiles
to the total mass density in each panel (gray, dashed lines); the scale height
of each fit is listed in the upper right of each panel.  Note that there is no
significant \s{0.0}{0.5} population at these annuli.}
\end{figure*}

We denote stellar populations grouped by age (age cohorts) with a script
$\mathcal{S}$ and subscripts corresponding to the population's range in
formation time; \eg, stars born between $0.0$ and $0.5$ Gyr after the big bang
will be collectively referred to as \s{0.0}{0.5}. Cohort age is the
difference between the age of the universe in the simulation ($13.327$
Gyr) and the cohort's formation time.

Figure~\ref{fig:400_den} shows the stellar surface density of the galaxy as a
function of position and age at redshift zero. The specific age cohorts are
defined by formation times (\tform) spanning $0.0$ to $0.5$ Gyr, $0.5$ to $1.0$
Gyr, $1.0$ to $2.0$ Gyr, $2.0$ to $4.0$ Gyr, $4.0$ to $8.0$ Gyr, $8.0$ to
$12.8$ Gyr, and the final $\sim500$ Myr of star formation, $12.8$ to $13.4$ Gyr
\footnote{The bin edge is at $13.4$ Gyr as it is the nearest $100$ Myr
increment that was inclusive of all stars.}.  Each panel denotes the range of
\tform\ considered and shows face-on and edge-on views of the galaxy. Following
the density distributions from left to right and top to bottom, younger stellar
populations show longer, thinner structure than older populations. The oldest
cohort (\s{0.0}{0.5}) is concentrated in the central region of the galaxy while
stars born during the next 500 Myr (\s{0.5}{1.0}) have a spheroidal mass
density profile out to a galactocentric radius (\rgc) of $\sim 10$ kpc.
Compared to stars with \tform$<1$ Gyr, \s{1.0}{2.0} extends further in radius,
and its edge-on view shows a distinct yet subdominant flattened component
indicative of an infant disk. \s{2.0}{4.0} is the first cohort to display
primarily non-spheroidal spatial structure; the edge-on view shows a vertically
extended and relatively compact disk. Edge-on views of \s{4.0}{8.0} and
\s{8.0}{12.8} reveal that younger populations inhabit progressively longer and
thinner disks.  \s{12.8}{13.4} is confined predominantly to the spiral arms and
has a very sharp break at $\rgc\approx10$ kpc (Figure~\ref{fig:400_den},
face-on view).  In time, the \s{12.8}{13.4} stars in these spiral arms will
heat up and increase their random motions, eventually leading to the
dissolution of the cohort's present-day structure (though newly formed stars
would enable the spiral arms to persist). The reduced azimuthal structure in
the density distributions of \s{4.0}{8.0} and \s{8.0}{12.8} are suggestive of
how the spiral overdensity can fade over time.  As descriptions of galaxy
morphology, the terms ``early'' and ``late'' have largely lost their
evolutionary connotations, but it is intriguing that the configurations of the
age cohorts within this individual simulated galaxy map almost perfectly onto
the Hubble sequence from early-type elliptical to late-type spiral.

\begin{figure}[h]
 \includegraphics[width=3.7in]{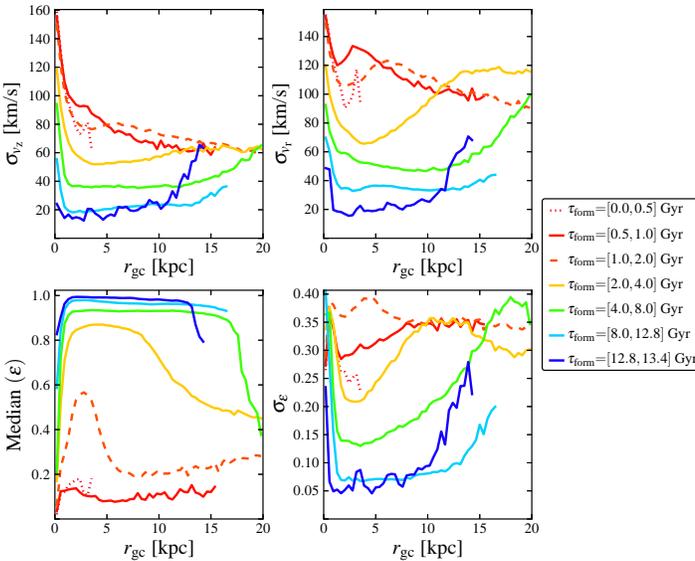}
 \caption{\label{fig:400_rp} The present-day radial profiles of vertical
 velocity dispersion (top left), radial velocity dispersion (top right), median
 circularity (bottom left, \cir$=\jz/\jcirc$, where \jz\ is a particle's angular momentum and \jcirc\ the angular momentum for a circular orbit with the particle's energy $E$), and dispersion of circularity (bottom right) for each
 age cohort. Younger cohorts are kinematically colder than older cohorts. The
 color of each radial profile indicates the cohort age as in
 Figure~\ref{fig:400_pos} (see legend).}
\end{figure} 

The age cohorts' surface mass density profiles  quantify their mass
contribution to the galaxy as a function of radius (Figure~\ref{fig:400_pos},
left panel). The color scheme for this figure is repeated throughout the rest
of the paper: red colors for the oldest cohorts, progressing through orange,
yellow, green, cyan, and dark blue for younger populations. A bulge-like
component dominates the inner 2 kpc of the surface mass density profile for all
stars (black line); starting at \rgc$\sim 3.5$ kpc, the profile becomes
exponential and shows no break out to $\rgc=20$ kpc. As seen in
Figure~\ref{fig:400_den}, \s{0.0}{0.5} contributes only to the central region
of the galaxy. The mass density profiles of  \s{0.5}{1.0} and \s{1.0}{2.0}
extend to larger radii and are intermediate in shape between a power law and
exponential characterization. In addition to a central power law distribution,
\s{2.0}{4.0} has a qualitatively exponential component at $3<\rgc<10$ kpc.
Younger populations show dominant exponential components: the mass density
profile of \s{4.0}{8.0} is exponential at $3<\rgc<20$ kpc while that of
\s{8.0}{12.8} has two exponential components over the same radial range with a
break at $\rgc\sim 12.5$ kpc. Stars born in the last $500$ Myr have the mass
profile of a exponential disk with a break radius at $9$ kpc. The cohorts also
show trends in their vertical structure: older populations are systematically
at greater heights above the plane than younger cohorts at all galactocentric
radii (right panel, Figure~\ref{fig:400_pos}). Several groups have measured a
similar correlation at the solar annulus in the MW and in external galaxies
\citep[\eg,][]{Bensby05, Yoachim08}. Note that the
precipitous rise in the median height above the plane in \s{12.8}{13.4} occurs
past the break radius of this population; only a few percent of this cohort's
mass is found in the disk outskirts, and these stars may be kinematically
distinct from those inside the break. The median height of all stars 
 is correlated with \rgc, partly because of an increasing halo to disk
ratio in the outskirts of the galaxy (see Section~\ref{sec:discussion} for the
same analysis applied to a kinematically selected disk sample). 

\begin{deluxetable}{llrrrrr}[ht]
\tablecolumns{7} 
\tablewidth{3.5 in} 
\tablecaption{Kinematic Decomposition} 
\tablehead{ 
  \colhead{\tform (Gyr)}    & \colhead{ID} & \colhead{$f_\mathrm{mass}$}   & \colhead{$f_\mathrm{thin}$} & \colhead{$f_\mathrm{thick}$} & \colhead{$f_\mathrm{pseudo}$} & \colhead{$f_\mathrm{spheroid}$}}
\startdata
(0.0,0.5)       & \s{0.0}{0.5}  &       $0.002$ &       $0.018$ &       $0.010$ &       $0.223$ &       $0.749$ \\
(0.5,1.0)       & \s{0.5}{1.0}  &        $0.058$ &       $0.021$ &       $0.053$ &       $0.182$ &       $0.743$ \\
(1.0,2.0)       & \s{1.0}{2.0}  &        $0.186$ &       $0.103$ &       $0.111$ &       $0.250$ &       $0.536$ \\
(2.0,4.0)       & \s{2.0}{4.0}  &        $0.313$ &       $0.438$ &       $0.143$ &       $0.211$ &       $0.208$ \\
(4.0,8.0)       & \s{4.0}{8.0}  &        $0.260$ &       $0.725$ &       $0.045$ &       $0.097$ &       $0.134$ \\
(8.0,12.8)      & \s{8.0}{12.8} &        $0.165$ &       $0.852$ &       $0.015$ &       $0.073$ &       $0.060$ \\
(12.8,13.4)     & \s{12.8}{13.4} &        $0.017$ &       $0.907$ &       $0.023$ &       $0.051$ &       $0.018$ \\
Total   &      &  $1.000$ &       $0.501$ &       $0.083$ &       $0.161$ &       $0.254$ \\
\enddata
\label{tab:decomp}

\tablecomments{The fractional contribution of each age cohort to different galactic structures at
redshift zero. The range of formation time, in Gyr, defining each age cohort is in
Column 1. The textual representation of each cohort is in Column 2. The fraction of present-day stellar mass in the cohort is in Column
3. The fraction of each cohort assigned to the thin disk, thick disk,
psuedobulge, and spheroid using our kinematic decomposition procedure (see
text) is listed in Columns 4, 5, 6, and 7, respectively.}
\end{deluxetable}

We show each age cohort's vertical density profile at three different
galactocentric radii in Figure~\ref{fig:400_vp}. The profile of all stars in
the solar annulus ($7<\rgc<9$ kpc, middle panel, black line) is consistent with the classic two
exponential profile shape observed in the MW \citep{Gilmore83} and recently
measured to have scale heights $z_{{ \mathrm h}_{thin}}=0.3$ kpc and  $z_{{
\mathrm h}_{thick}}=0.9$ kpc \citep{Juric_etal08}; the simulated galaxy's scale
heights are a factor of $1.6$-$1.8$ larger at $z_{{\mathrm h}_1}=0.55$ kpc and $z_{{\mathrm
h}_2}=1.47$ kpc but maintain the same thick to thin disk ratio.  Traditionally,
this double-exponential vertical density profile has been interpreted as two
distinct structural components, but Figure~\ref{fig:400_vp} clearly shows that
the superposition of all cohort vertical density profiles, spanning a continuum
of scale heights (increasing with age), naturally yields the double-exponential
profile (see Section~\ref{sec:discussion}).  Incorporating data from the inner
and outer disk (left and right panels, respectively), we find two general
trends governing the relationship between age, scale height, and radius: older
stellar populations have more extended vertical density profiles than younger
age cohorts, while each individual cohort's vertical profile becomes more
extended at larger radii.

The galaxy has dynamical trends with age as well: the orbits of older age
cohorts are systematically hotter than their younger counterparts in the radial
and vertical directions at nearly all radii (top row of
Figure~\ref{fig:400_rp}). The same general trend is seen in studies based on the Geneva Copenhagen Survey in the MW \citep{Holmberg07, Casagrande11}, though some explorations of smaller data sets find a plateau in the age-velocity relationship for stars of intermediate age
\citep[\eg,][]{Wielen77,Quillen01,Soubiran08}. Each individual age cohort's
velocity dispersion profiles have note-worthy features. The velocity
dispersions of \s{0.5}{1.0} and \s{1.0}{2.0} are locally maximal at the edges
of their bulge-like mass distributions ($\rgc \sim 3$ and $6$ kpc,
respectively). Each cohort of stars with \tform$>4$ Gyr shows little
internal variation in velocity dispersion as a function of radius in the galaxy's
disk out to \rgc$\sim 12.5$ kpc; thereafter, there is a positive gradient in
\vrdisp\ and \vzdisp\ with radius. Interestingly, the mass density profile of
\s{2.0}{4.0} is consistent with that of an exponential disk at \rgc$>5$ kpc
(Figure~\ref{fig:400_rp}), yet this radius is precisely the starting point for
a large, positive radial gradient in its \vrdisp\ and (to a lesser degree)
\vzdisp\ profiles, which is not seen in the younger cohorts except for the
sparse outskirts of \s{12.8}{13.4}.  It is plausible that the vertically
extended disk of \s{2.0}{4.0} experienced a significantly different
evolutionary history than the younger disk populations.

The bottom row of Figure~\ref{fig:400_rp} examines the orbital shape
(circularity) of cohort members as a function of radius. To define orbital
circularity (\cir), we first choose a coordinate system such that the origin is
coincident with the stellar particles' center of mass and the $z$-axis is
aligned with the total angular momentum vector of the stars. Each particle's
angular momentum in the plane of the disk is then \jz, and positive \jz\
denotes a corotating orbit. Circular orbits have the maximal \jz\ (hereafter
\jcirc) given a particle's total energy ($E_\mathrm{tot}$, see below).  A
particle's circularity (\cir) is defined as \jz$/$\jcirc$(E_\mathrm{tot})$.
Following the trends in velocity dispersion, older stars are on less circular
orbits and show more variation in their orbital shape than younger populations.
The median circularity within the disk components of \s{2.0}{4.0},
\s{4.0}{8.0}, and \s{8.0}{12.8} remains constant over a radial range that
increases when younger populations are considered. The dispersion in
circularity, however, grows with radius for these three cohorts; the radial
onset of the increase is inversely correlated with age. Stars formed during the
last $500$ Myr are generally born on very circular orbits and only have
significant circularity dispersion past the \s{12.8}{13.4} disk break radius.

\begin{figure}[!t] 
  \includegraphics[width=3.7in]{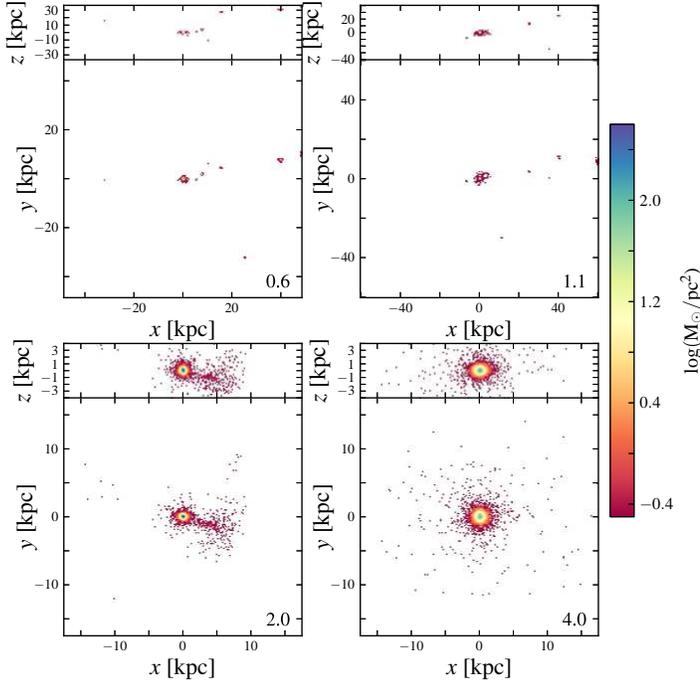}
  \caption{\label{fig:tf0.0_den} The surface density of \s{0.0}{0.5} as a
  function of position and time. Edge-on and face-on views are shown at the
  time of each snapshot (labeled in gigayears at the bottom right hand corner
  of each panel). Pixel color represents the logarithmic surface mass density
  at the pixel position (see colorbar). The coordinate system is such that the
  disk lies in the $xy$ plane and the galaxy's angular momentum vector is
  aligned with the $z$ axis.  Note the changing spatial scale: for clarity, our
  axis limits either correspond to the box encompassing $95\%$ of the cohort's
  mass or the region considered in Figure~\ref{fig:400_den} ($|x|,|y|,|z| <
  17.5,17.5,4.0$ kpc), whichever is larger. \s{0.0}{0.5} assembles in the
  center of the galaxy quickly. Snapshots at later times show no qualitative
differences with $t=4.0$ Gyr.} 
\end{figure} 

\begin{figure*}[!ht]
  \includegraphics[width=\textwidth]{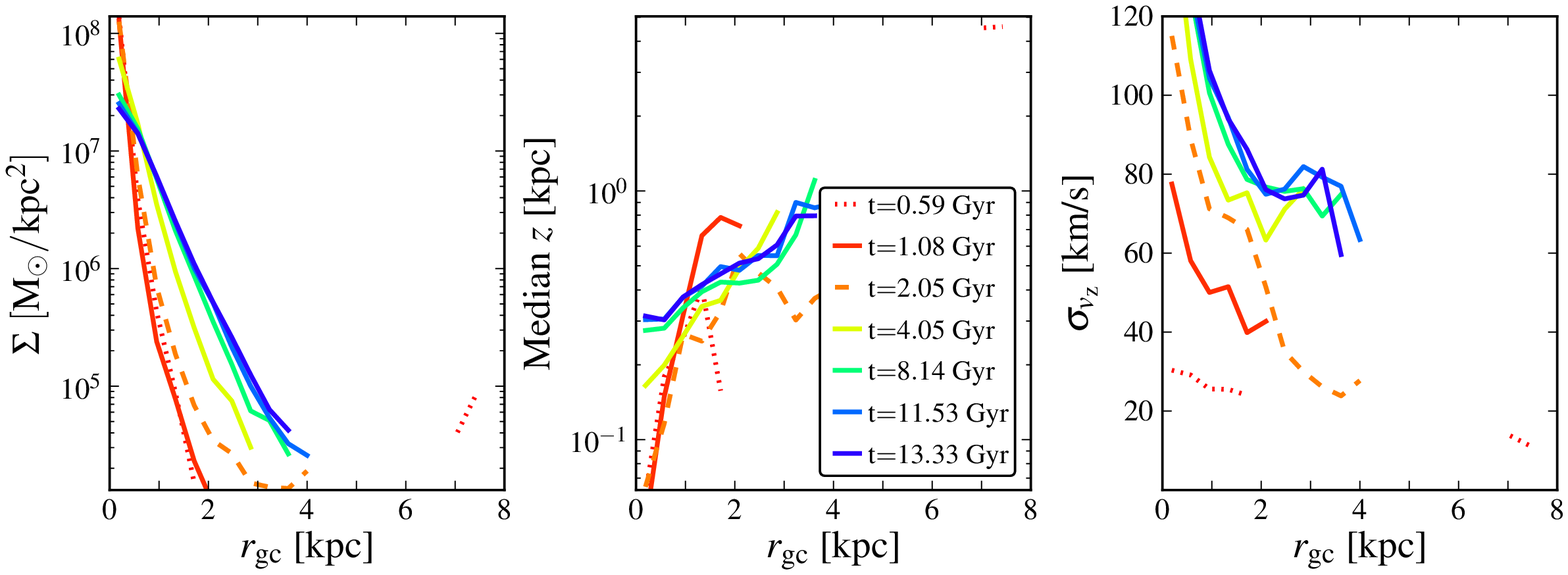}
  \caption{\label{fig:tf0.0_rp} The radial profiles of surface mass density
  (left panel), median height above the disk plane (middle panel), and vertical
  velocity dispersion (right panel) for \s{0.0}{0.5} as a function of time. The
color of each line represents the age of the simulation when the profile was
measured (see legend in middle panel). Note that the colors progress from red
for early times to blue for late times; we adopt this color scheme throughout
the rest of the figures. Line types help differentiate the earliest outputs.}
\end{figure*} 

To aid in our description of the current state of the galaxy, we kinematically
separate the galaxy into structural components using particle energy and
angular momentum and following a modified version of the prescription outlined
in \citet{Abadi03}. We adhere to convention and describe these components as
the ``thin disk'', ``pseudobulge'', \etc, though this nomenclature implies, to
some degree, sharp distinctions in the observed continuous distribution.

The principal discriminants in our decomposition are circularity and total
specific energy defined by $E_\mathrm{tot}=\frac{1}{2}v^2 + \Phi$; where $v$ is
the total velocity of the particle and $\Phi$ is the value of the potential (an
output of the simulation code) at the position of the particle. We first assign
all particles with $\cir>0.8$ to the thin disk as it should contain mostly
circular orbits ($\cir=1$ for a circular orbit).  We assign to the spheroid all
particles with circularity below a critical threshold $\jcrit=0.33$, which is
chosen such that the ensemble of all particles with \cir$<\jcrit$ exhibit zero
net rotation (all counter-rotating particles are assigned to the spheroid). The
remaining population has \jcrit$<\cir<0.8$ and is subdivided into the thick
disk ($\Etot\ge E_\star$) and pseudobulge ($\Etot< E_\star$) based on the
median energy of all stellar particles ($E_\star$). Note that this definition
of pseudobulge is not identical to that used in \citet{Guedes12}, where the
pseudobulge was first identified photometrically as the stellar mass contained
in the inner $2$ kpc, which has a higher Sec profile index than the disk.
Their pseudobulge was then constrained further by including only stars with
circularity below $0.8$ to exclude disk stars. A fraction of stars that we
identify as belonging to the thick disk via the energy cut or to the spheroid
via the circularity threshold would have been classified as pseudobulge using
the \citet{Guedes12} definition. 

Our kinematic distinctions are somewhat arbitrary and may not perfectly
describe the galaxy. The caveats to this decomposition are similar to those in
\citet{Abadi03}: the spheroidal component could have a modest amount of
rotation in reality, the circularity cutoff for the thin disk may not precisely
match that in the MW, and the energy cuts do not ensure that our structural
distinctions are complete and free of contamination. In addition, our choice to
uniquely assign all regions of the $E_\mathrm{tot}$, \cir\ plane to specific
structures demands that the \cir\ distribution of the spheroid is not symmetric
about \cir$=0$, extending to \cir$=-1$ for counter-rotating stars but stopping
at \jcrit\ for co-rotating orbits. We stress that our subsequent analysis and
conclusions are not based on these kinematic selections.  However, the
decomposition gives us a qualitative description of the galaxy's kinematic
structure that provides a familiar context for our analysis.

In Table~\ref{tab:decomp}, we show how each age cohort contributes to the
galactic structures identified in our kinematic decomposition. The results
corroborate much of what we see in Figure~\ref{fig:400_den}. Old stellar
populations with \tform$<2.0$ Gyr predominantly contribute to the spheroid.
After $t=4$ Gyr, new stars overwhelmingly populate the disk of the galaxy.
\s{2.0}{4.0} is an intermediate population: significant fractions of its
members have kinematics consistent with classical disk and spheroid orbits. The
total fractions confirm that the simulated galaxy has a strong disk
component.  The majority of all particles have high circularity
($f_{\cir>0.7}=0.6$), similarly prominent high circularity populations are 
found in the most disk-dominated galaxies simulated with the latest
implementation of the Tree-SPH code GADGET-3 \citep{Aumer13}.  

\begin{figure}[t]
  \includegraphics[width=3.7in]{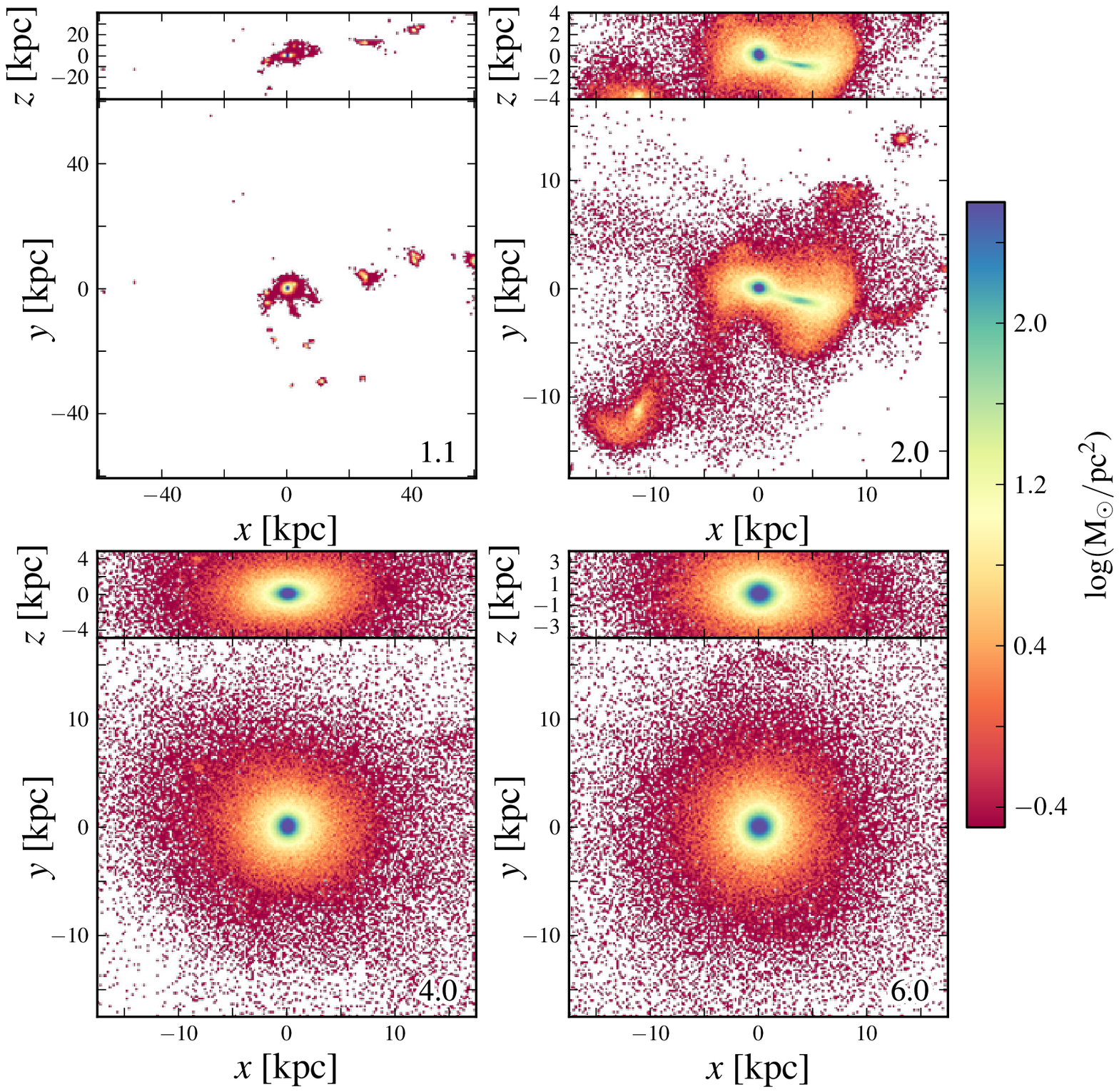}
  \caption{\label{fig:tf0.5_den} The surface density of \s{0.5}{1.0} as a
  function of position and time. The orientation of the galaxy, calculation of axis limits,
  and color scheme are the same as in Figure~\ref{fig:tf0.0_den}. The time of
  each snapshot is labeled in gigayears at the bottom right hand corner of each
panel. Later outputs are omitted as they show no qualitative changes since
$t=6.0$ Gyr.} \end{figure}   

\section{Evolution of Age Cohorts}
\label{sec:evolution}

\begin{figure*} 
  \includegraphics[width=\textwidth]{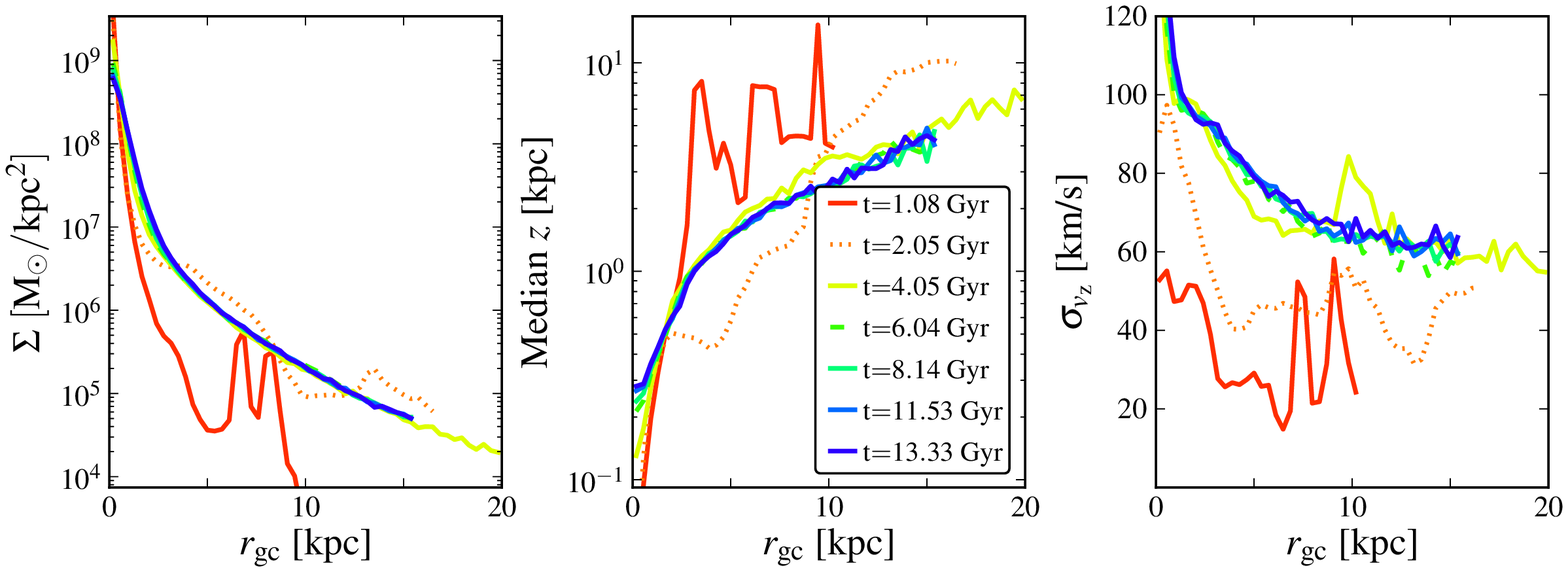}
  \caption{\label{fig:tf0.5_rp} The radial profiles of surface mass density
  (left panel), median height above the disk plane (middle panel), and vertical
  velocity dispersion (right panel) for \s{0.5}{1.0} as a function of time. As
  in Figure~\ref{fig:tf0.0_rp}, line color and type represent the age of the
  simulation when the profile was measured. The profiles now extend to $\rgc=20$ kpc.}
\end{figure*}  

We now examine the assembly history of the galaxy using the individual age
cohorts as tracer populations. 
In this section we adjust our choice of age cohorts so that each represents
at most a 1 Gyr range of formation times.
The first three \tform\ bins
(\s{0.0}{0.5}, \s{0.5}{1.0}, and  \s{1.0}{2.0}) are illustrative of the
relatively complex assembly of stellar populations at early times.
\s{3.0}{4.0} is an ``intermediate'' population born just after the major merger
epoch concludes. The final age cohort described here, \s{7.0}{8.0}, is 
representative of populations born and evolved {\it in situ} in a quiescent 
disk.

\begin{figure}[t]  
  \includegraphics[width=3.7in]{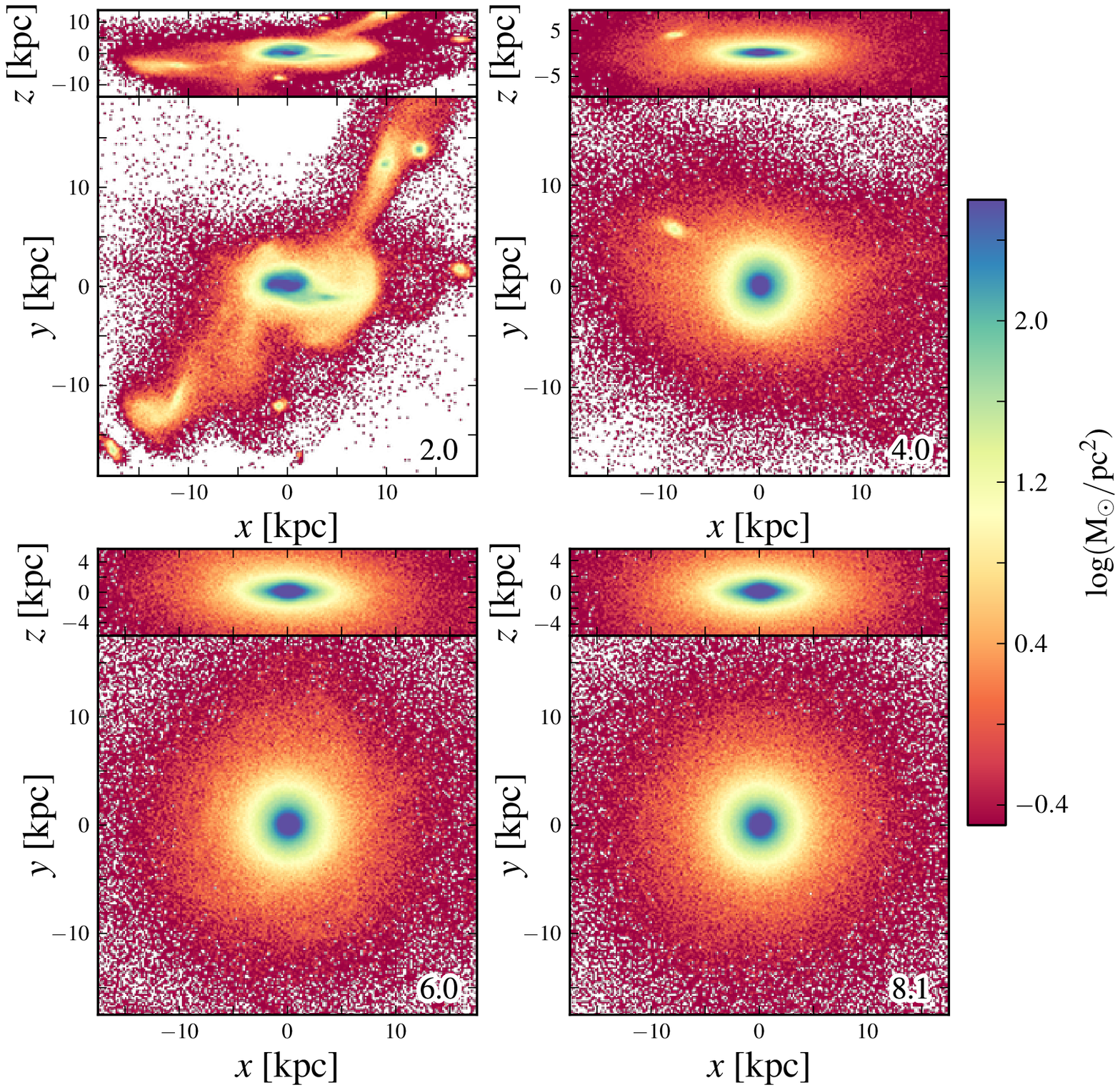}
  \caption{\label{fig:tf1.0_den}  The surface density of \s{1.0}{2.0} as a
  function of position and time. The orientation of the galaxy, calculation of axis limits,
  and color scheme are the same as in Figure~\ref{fig:tf0.0_den}. The time of
  each snapshot is labeled in gigayears at the bottom right hand corner of each
  panel. Later outputs are omitted as they show no qualitative changes since
$t=8.1$ Gyr.}
\end{figure} 
 
\textbf{The First 500 Myr} We consider the surface density distribution of
\s{0.0}{0.5} as a function of time, specifically at $t = 0.6$, $1.1$, $2.0$,
and $4.0$ Gyr, in Figure~\ref{fig:tf0.0_den}. These stars populate a number of
different subhalos at early times, but their dominant concentration is in the
parent halo of the galaxy (note the changing scale on each panel in the
figure). The final merger event involving a perceptible fraction of
\s{0.0}{0.5} is well underway at $t=2$ Gyr; by $t=4$ Gyr, \s{0.0}{0.5} has
assembled into a configuration much like what we see today (compare the bottom
right of Figure~\ref{fig:tf0.0_den} with the top left of
Figure~\ref{fig:400_den}). We omit snapshots at later times as \s{0.0}{0.5}
shows no further qualitative changes in its density distribution. 

\begin{figure*} 
  \includegraphics[width=\textwidth]{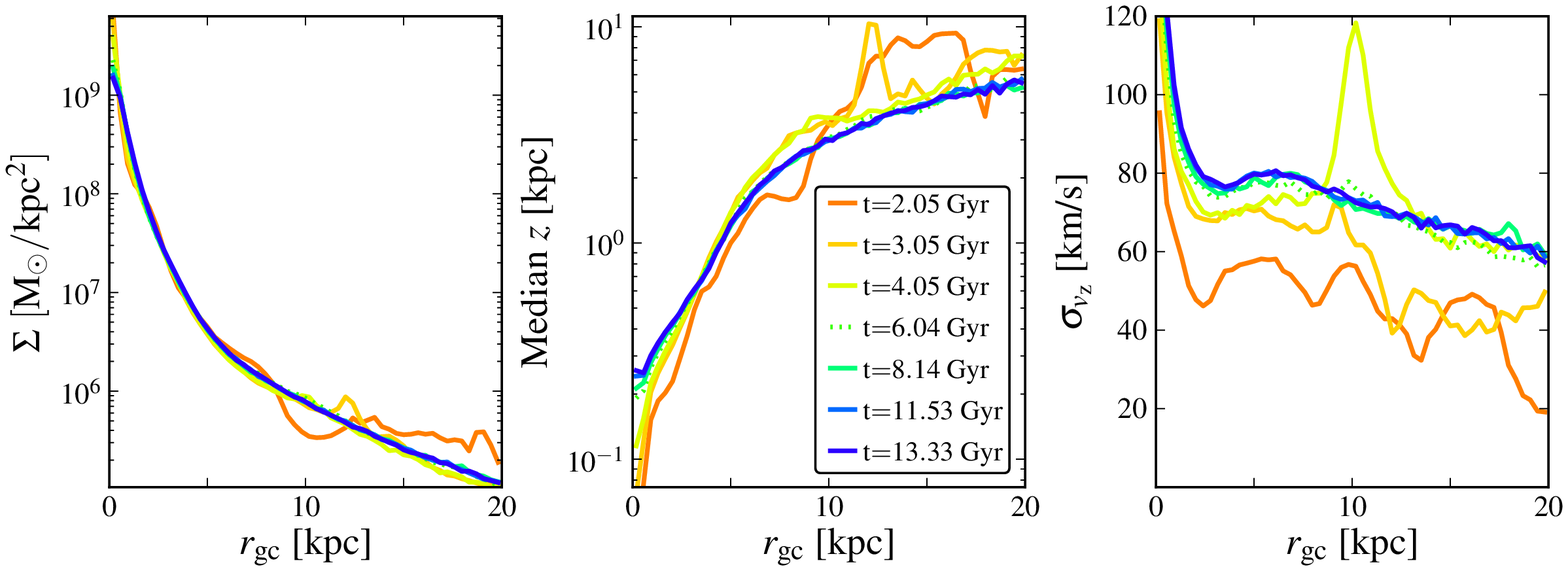}
  \caption{\label{fig:tf1.0_rp} The radial profiles of surface mass density
  (left panel), median height above the disk plane (middle panel), and vertical
  velocity dispersion (right panel) for \s{1.0}{2.0} as a function of time. As
  in Figure~\ref{fig:tf0.0_rp}, line color and type represent the age of the
simulation when the profile was measured.}
\end{figure*} 

Physical parameters describing spatial structure and kinematics confirm that
the \s{0.0}{0.5} population assembles fairly rapidly and evolves quiescently
thereafter (Figure~\ref{fig:tf0.0_rp}). Both the satellite accretion and the
decreasing central density seen in Figure~\ref{fig:tf0.0_den} broaden the
cohort's radial surface mass density profile with time. The radial profiles of
vertical velocity dispersion (right panel) and median height above the plane
(middle panel) show that \s{0.0}{0.5} kinematically heats up as it assembles
but only makes moderate energy gains after $t=4$ Gyr. \s{0.0}{0.5} congregates
early in the galaxy's history, remains centrally concentrated, and evolves
quiescently after its last substantial merger event.

\textbf{$\mathbf{ \tform=[0.5,1.0]}$ Gyr} Similar to \s{0.0}{0.5}, \s{0.5}{1.0}
is scattered amongst several subhalos at early times but most of its mass is
still born in the parent halo (Figure~\ref{fig:tf0.5_den}). The merger event at
$t=2$ Gyr is more significant for \s{0.5}{1.0} than \s{0.0}{0.5}. Still, after
another $2$ Gyr, the morphological features associated with the encounter have
dissipated.  From $t=4$ Gyr to $t=6$ Gyr, the density distribution of stars
outside \rgc$>6$ kpc becomes more symmetric. Later snapshots do not show any
qualitative changes.

The surface mass density radial profile of \s{0.5}{1.0} resembles its current
state throughout most of the galaxy by $t\sim4$ Gyr; however, a perceptible
decrease in  mass within the innermost kiloparsec and corresponding increase in $1<\rgc<3$ kpc persists until $t=8$ Gyr (Figure~\ref{fig:tf0.5_rp}, left panel). \s{0.5}{1.0} is born
with more random vertical motion than \s{0.0}{0.5} initially had; like
\s{0.0}{0.5}, \s{0.5}{1.0} also experiences an early, dramatic increase in its
vertical energy (Figure~\ref{fig:tf0.5_rp}, right panel).  Like \s{0.0}{0.5},
\s{0.5}{1.0} shows almost no evolution of the vertical velocity dispersion
after $t=4$ Gyr.  Only the central kiloparsec shows any appreciable change in
the median height of the cohort after this time, with growth of the median
height by about 100 pc (Figure~\ref{fig:tf0.5_rp}, middle panel). \s{0.5}{1.0},
like \s{0.0}{0.5}, quickly becomes the kinematically hot, vertically extended
population seen at redshift zero (Section~\ref{sec:current}). We also note that
a central bulge-like component with steeper density profile slope has already
formed at early times as a result of bar-like instabilities in the early disk
\citep[see][]{Guedes12}.

\textbf{$\mathbf{ \tform=[1.0,2.0]}$ Gyr} The surface density distribution of
\s{1.0}{2.0} is shown as a function of time in Figure~\ref{fig:tf1.0_den}.
\s{1.0}{2.0} has a significant {\it in situ} population that forms either 
during or
just prior to the merger event at $t=2.0$ Gyr. This merger strongly impacts the
cohort's morphology. In spite of these turbulent beginnings, almost all
of the \s{1.0}{2.0} stars have been assimilated into a coherent structure by
$t=4$ Gyr. \s{1.0}{2.0}'s configuration is the first to show an ellipsoidal
disk component; the disk is evident just after the cohort forms ($t=2$ Gyr) and
is obvious by $t=4$ Gyr (see edge-on views). From $t=4$-$8$ Gyr, the density
distribution at \rgc$>5$ kpc becomes more smooth, and any discernible tilt to 
the ellipsoid vanishes.

Despite differences in their assembly history, \s{1.0}{2.0}'s kinematic
properties are similar to those of older populations. The cohort's radial
surface density profile remains constant after $t=4$ Gyr, except for some
emigration away from the innermost kiloparsec (Figure~\ref{fig:tf1.0_rp}, left
panel). The initial, {\it in situ} population is born with relatively large \vzdisp,
yet there is still a substantial increase in the random vertical motions of
this cohort as its satellite-born populations merge (Figure~\ref{fig:tf1.0_rp},
right panel), including the satellite still visible in the $t=4$ Gyr snapshot.
As with older cohorts, there is little heating at late times, and
both the median \vzdisp\ and median height profiles show essentially
no evolution after $t=6$ Gyr.

\begin{figure}[t]
\includegraphics[width=3.7in]{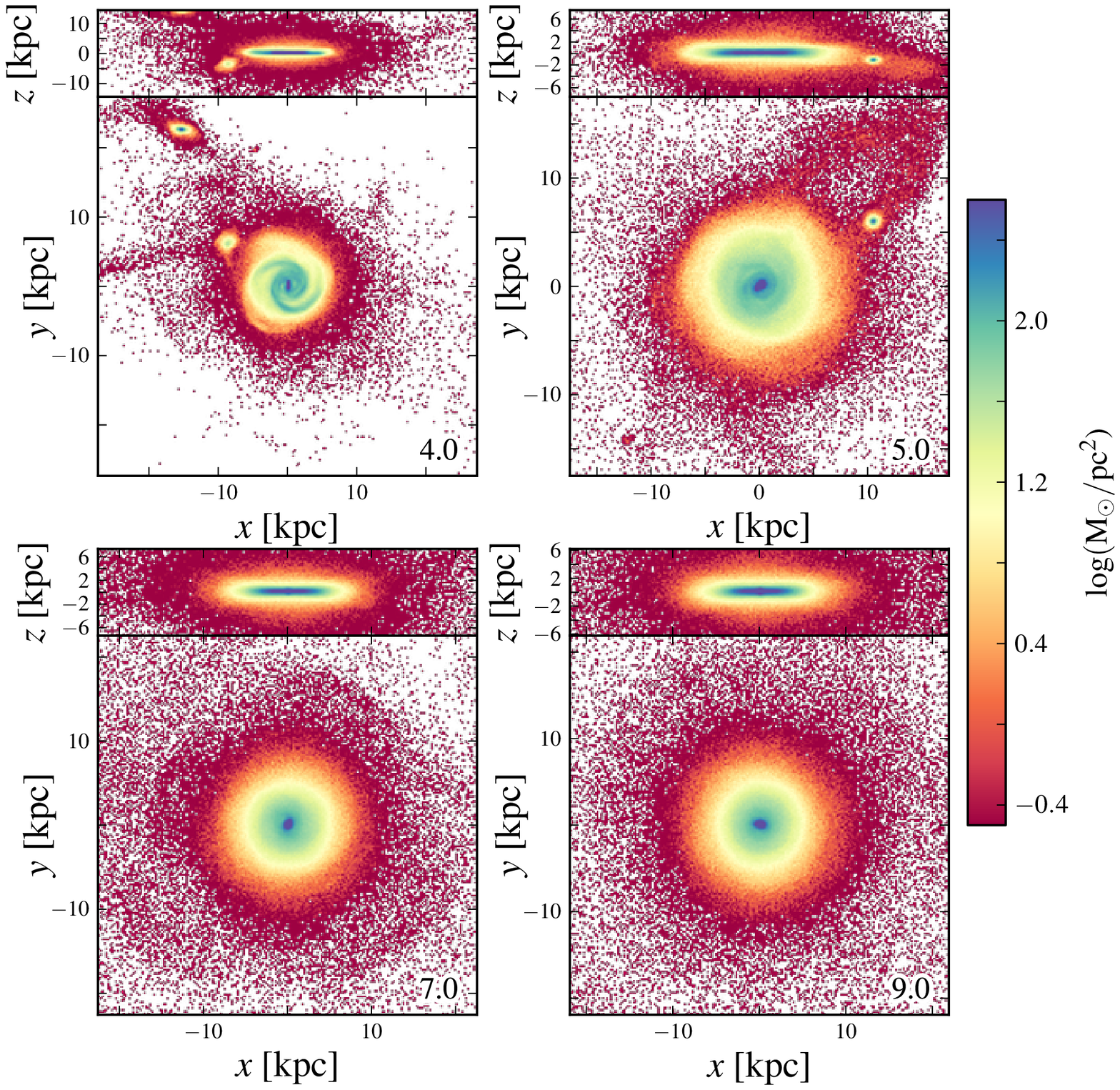}
\caption{\label{fig:tf3.0_den} The surface density of \s{3.0}{4.0} as a
function of position and time. The orientation of the galaxy, calculation of axis limits,
and color scheme are the same as in Figure~\ref{fig:tf0.0_den}. The time of
each snapshot is labeled in gigayears at the bottom right hand corner of each
panel. Later outputs are omitted as they show no qualitative changes since
$t=9.0$ Gyr.}
\end{figure}  

\begin{figure*}
\includegraphics[width=\textwidth]{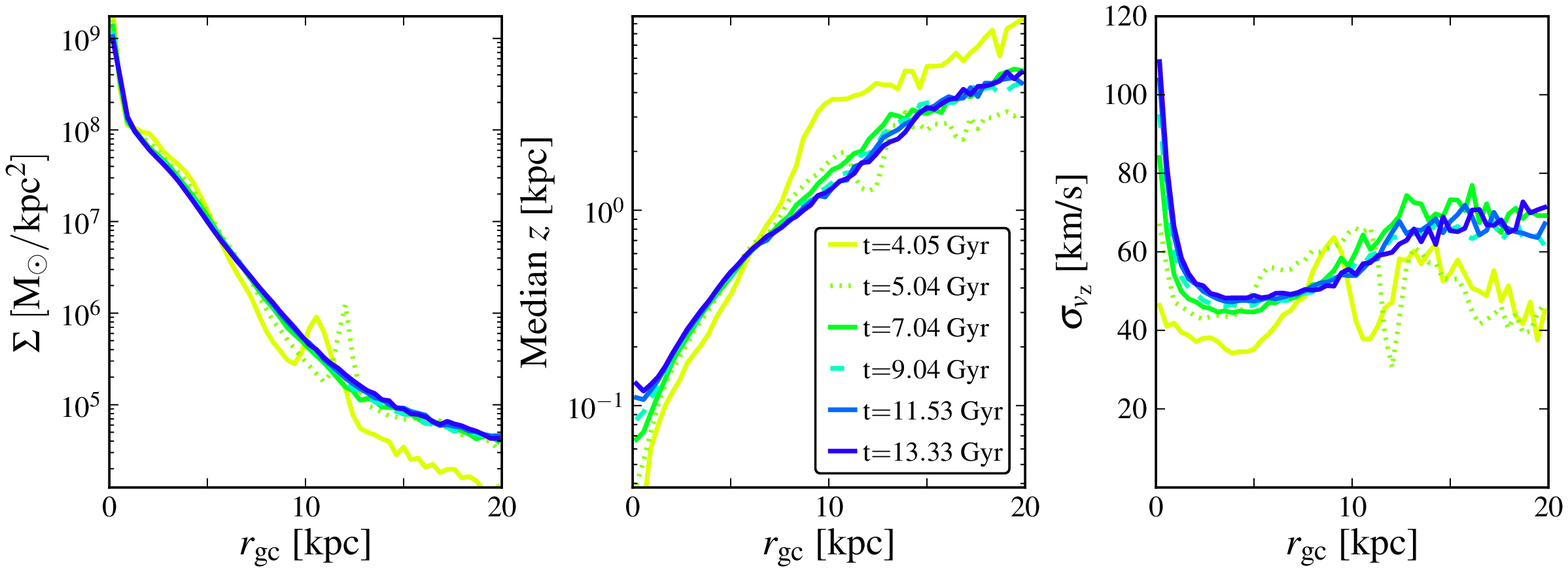}
\caption{\label{fig:tf3.0_rp}  The radial profiles of surface mass density
(left panel), median height above the disk plane (middle panel), and vertical
velocity dispersion (right panel) for \s{3.0}{4.0} as a function of time. As in
Figure~\ref{fig:tf0.0_rp}, line color and type represent the age of the
simulation when the profile was measured.}
\end{figure*} 

\textbf{$\mathbf{ \tform=[3.0,4.0]}$ Gyr} Many of the evolutionary trends
established for stars with \tform$< 2$ Gyr are no longer evident when we examine
\s{3.0}{4.0} (Figure~\ref{fig:tf3.0_den}). Here, the vast majority of the
cohort is born {\it in situ} in the galactic disk; the early \s{3.0}{4.0} disk shows
a prominent $m=2$ mode spiral (top left of Figure~\ref{fig:tf3.0_den}). The
initial snapshot also reveals two significant satellites that will soon
interact with the {\it in situ} population. At $t=5$ Gyr (top right), one of the two
satellites is fully disrupted and the cohort's strong spiral structure has
dissipated. Over the next two billion years, the second satellite merges, the
spiral structure dissolves, and there is a small but perceptible
thickening of the disk. Stars at large radii continue to phase-mix over time
($t=9$ Gyr, bottom right). 

\s{3.0}{4.0} is the oldest cohort to have a dominant exponential component in
its surface density radial profile (Figure~\ref{fig:tf3.0_rp}, left panel).  As
seen in the density plots, this cohort's disk was in place at early times.  The
density profile shows some mass redistribution from the interior of the disk
(\rgc$<6$ kpc) to the outer disk over time, moderately increasing the cohort's
disk scale length. We investigate radial migration patterns of this cohort and
others in Section~\ref{sec:radmix_tf}. The radial profile of \vzdisp\ (right
panel) shows that \s{3.0}{4.0}, while kinematically hot for a disk population,
is born with lower internal energy than the older cohorts.  \s{3.0}{4.0}
members within \rgc$<6$ kpc become modestly more energetic with time
(Figure~\ref{fig:tf3.0_rp}, right panel); over the last $10$ Gyr, the
population's heating rate ($\Delta\vzdisp / \Delta t$) is twice that of
\s{1.0}{2.0}. While the older populations became colder at increasing radius,
\s{3.0}{4.0} has a positive \vzdisp\ radial gradient; we suspect that this is
most likely due to relatively cold, disk orbits at smaller radii giving way to
hotter halo kinematics past the break radius of the \s{3.0}{4.0} disk (Section~\ref{sec:discussion} examines a kinematically selected disk sample).  
The median height of the cohort increases between $t=4$ and $t=5$ Gyr
at \rgc$<5$ kpc, while the outer disk settles to a thinner configuration.

\begin{figure}[t]
\includegraphics[width=3.7in]{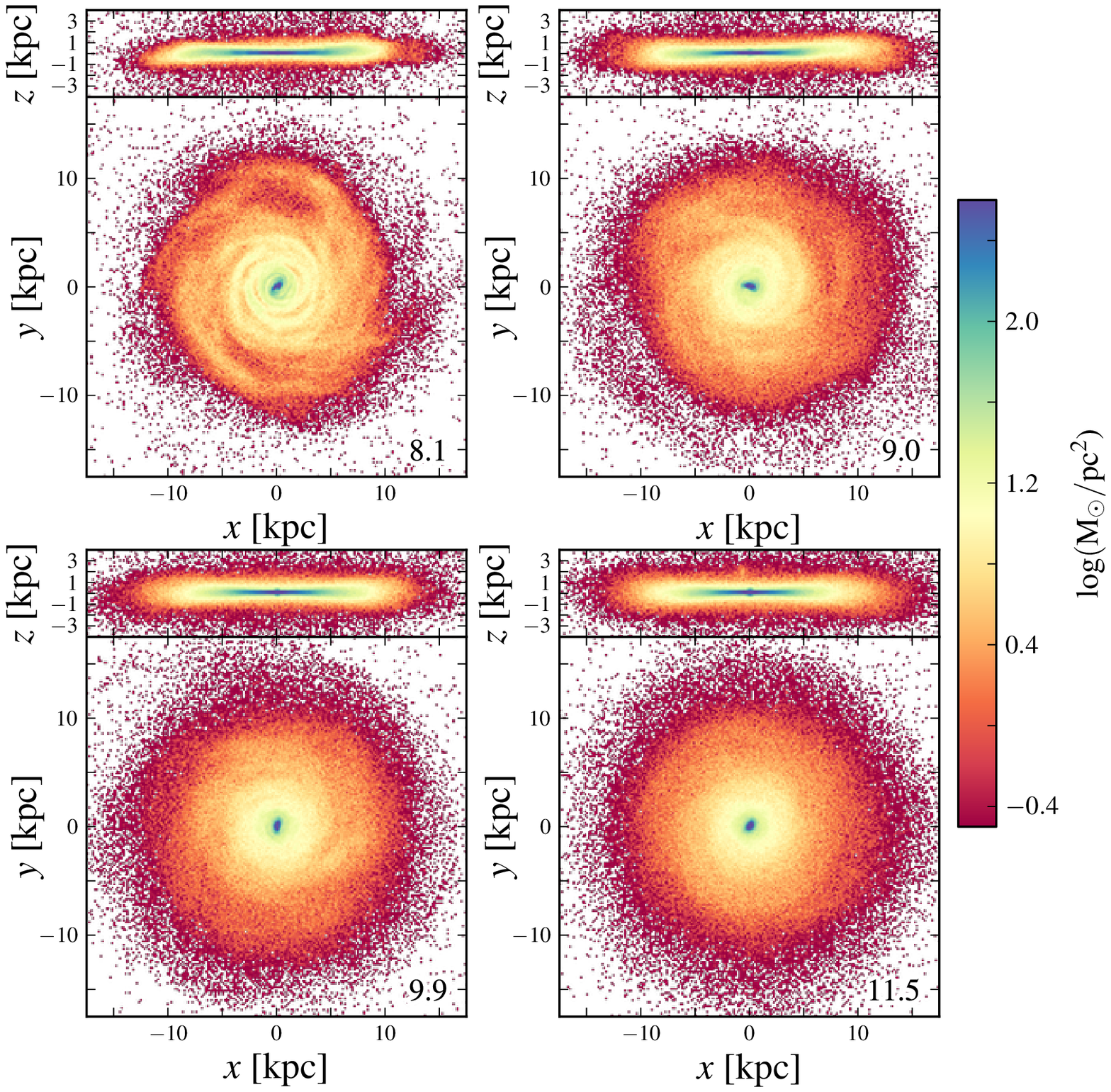}
\caption{\label{fig:tf7.0_den} The surface density of \s{7.0}{8.0} as a
function of position and time. The orientation of the galaxy, calculation of axis limits,
and color scheme are the same as in Figure~\ref{fig:tf0.0_den}. The time of
each snapshot is labeled in gigayears at the bottom right hand corner of each
panel. Later outputs are omitted as they show no qualitative changes since
$t=11.5$ Gyr.}
 \end{figure} 

\begin{figure*}[!ht] 
\includegraphics[width=\textwidth]{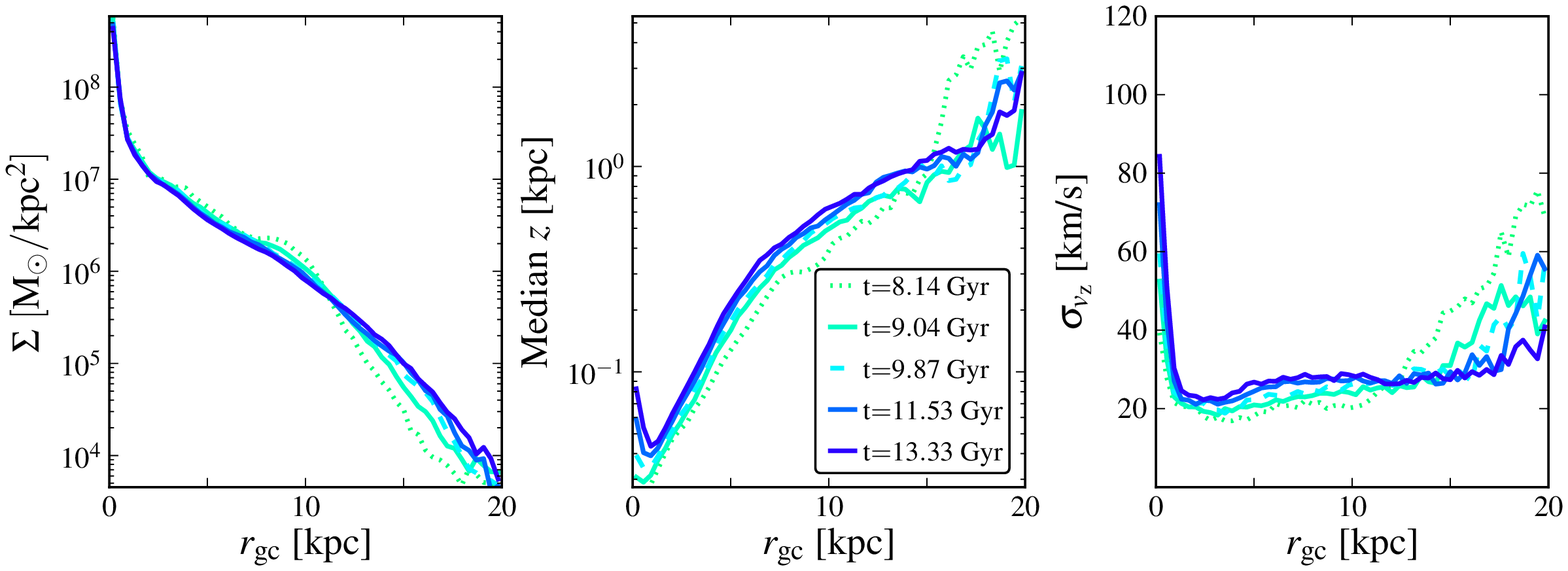}
\caption{\label{fig:tf7.0_rp} The radial profiles of surface mass density
(left panel), median height above the disk plane (middle panel), and vertical
velocity dispersion (right panel) for \s{7.0}{8.0} as a function of time. As in
Figure~\ref{fig:tf0.0_rp}, line color and type represent the age of the
simulation when the profile was measured.}
\end{figure*} 

\textbf{$\mathbf{ \tform=[7.0,8.0]}$ Gyr} \s{7.0}{8.0} is illustrative of
stellar populations born after redshift one: it forms almost exclusively in the
disk and secularly evolves (Figure~\ref{fig:tf7.0_den}). The initial disk of
\s{7.0}{8.0} is thin and shows obvious warping in its outskirts
(Figure~\ref{fig:tf7.0_den}, top left panel). During this cohort's evolution, a
small bulge-like component re-grows in the central region of the galaxy
\citep[see][]{Guedes12}, the initial outer disk warps weaken, and the cohort's
disk becomes more vertically and radially extended.

The cohort's surface density radial profile clearly shows a central component
giving way to an exponential disk at \rgc$\sim 2$ kpc
(Figure~\ref{fig:tf7.0_rp}, left panel). Like \s{3.0}{4.0}, there is a net
transfer of mass outwards; here, the effect is more dramatic and the donor
region extends out to $\rgc\sim11$ kpc. 
The inner disk (\rgc$\la 10$ kpc), born cold, experiences significant vertical
heating, with \vzdisp\ at the solar radius growing from 
$\approx 20$ \kms\ to $\approx 30$ \kms\
(Figure~\ref{fig:tf7.0_rp}, right panel).
More cohort members populate the sparse outer disk, beyond \rgc$\approx 12$ kpc, over time and reduce the vertical velocity dispersion there. 
The central mass concentration seems distinct from the
disk; stars in the central few hundred parsecs continue to move further from
the plane throughout the cohort's history. In a departure from previously
examined cohorts, \s{7.0}{8.0}'s entire disk slowly grows more vertically
extended as a function of time (middle panel). The gradual increases
in vertical velocity dispersion and median particle height are fractionally
large compared to the late-time evolution of older cohorts, but they are still
modest on an absolute scale.

\section{Radial Migration}
\label{sec:radmix_tf}

Radial migration, \ie, stellar orbital angular momentum changes brought about by resonant interactions, can have a profound impact on the evolution of disk galaxies
\citep[][and references therein]{Schonrich09, Roskar12a}. We now investigate
the age cohorts' migratory patterns. Stars are grouped into the same age
cohorts used in Section~\ref{sec:current}, except that all stars with \tform$<2$
Gyr are now labeled as a single cohort. We limit contamination from halo
stars by constraining our
migration analysis to particles within $4$ kpc of the galactic plane at $z=0$
\footnote{Relatively few particles have heights $z>4$ kpc; our main results are
independent of this selection cut.}. 

\begin{figure*}[p]
\includegraphics[width=\textwidth]{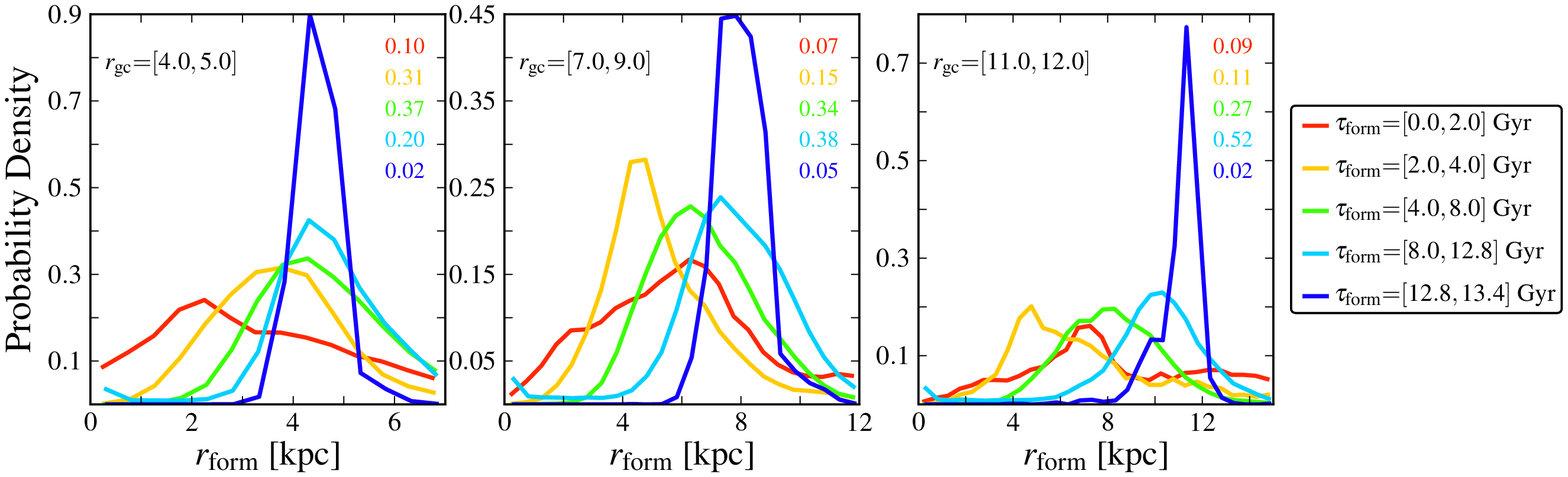}
\caption{\label{fig:rform}The distribution of formation radius for stars in
different age cohorts residing at $\rgc=[4.0,5.0]$, $\rgc=[7.0,9.0]$, and
$\rgc=[11.0,12.0]$ at redshift zero (left to right, respectively). Color-coding differentiates the age cohorts and refers to their
formation time (see legend). The fraction of mass
within each annulus associated with each cohort is labeled with the
corresponding color in the top right corner of each panel (listing is
chronological by formation time). }
\end{figure*}

Figure~\ref{fig:rform}  shows the birth
radii for cohort members currently residing in three different areas of the
disk.  The middle panel examines the present day solar annulus ($7\leq\rgc\leq9$
kpc). The oldest inhabitants of the solar annulus (\s{0.0}{0.2}) come from the
broadest range of formation sites.  Stars born at large radii likely formed in
a satellite halo before merging with the {\it in situ} population.  Members of
\s{0.0}{2.0} are predominantly associated with the spheroid
(Section~\ref{sec:current}) and only contribute $7\%$ of the total mass at this
annulus (see Figure legend). For current solar annulus stars with \tform$>2$
Gyr, progressively younger stars are less centrally concentrated at birth; the
median \rform\ is $5.2$, $6.6$, $7.8$, and $7.9$ kpc for \s{2.0}{4.0},
\s{4.0}{8.0}, \s{8.0}{12.8}, and \s{12.8}{13.4}, respectively. Inside-out
growth ensures that the same correlation between age and formation radius found
in the solar annulus exists  both in the inner (left panel) and outer (right
panel) disk. Excluding \s{0.0}{2.0} and young stars with little time to migrate
(\s{12.8}{13.4}), the median radial excursion ($r_{\mathrm{final}}-$\rform) of
a cohort increases with cohort age and the final radius considered. Though the
galaxy was built up in an inside-out fashion, it is important to note that the
birth sites of old stars are not restricted to the central galaxy.  For example,
$20\%$ of \s{2.0}{4.0} members currently in the solar annulus have
$6<\rform<10$ kpc.

\begin{figure*}[p]
\includegraphics[width=\textwidth]{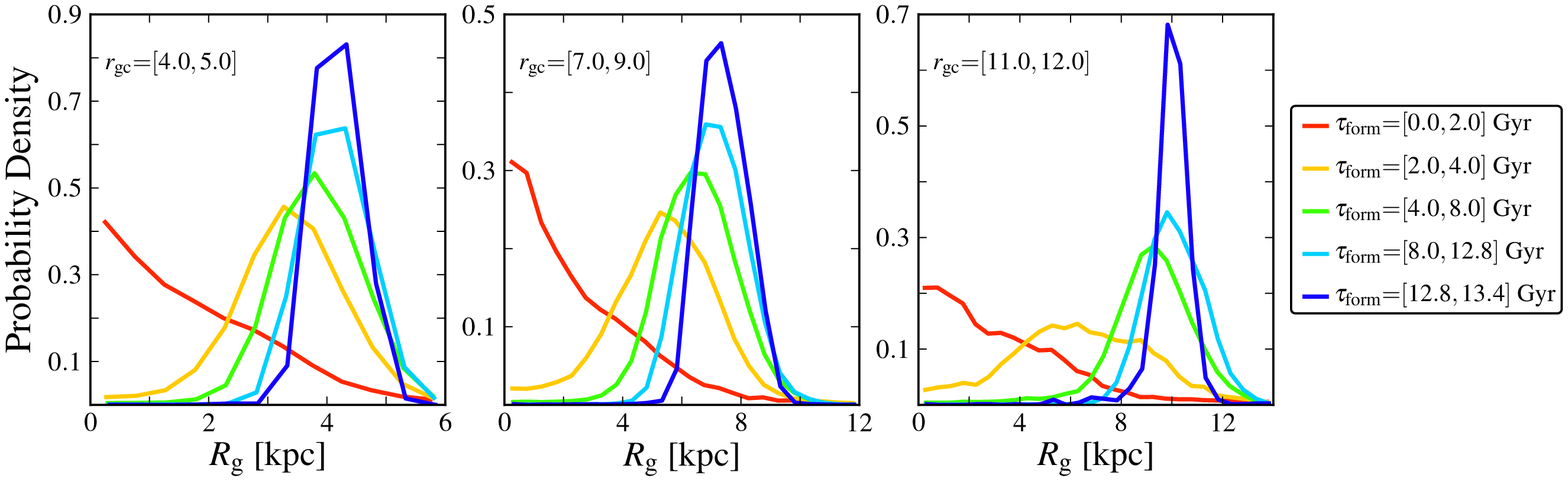}
\caption{\label{fig:gcr}The current guiding center radius of age cohort members
residing at $\rgc=[4.0,5.0]$, $\rgc=[7.0,9.0]$, and $\rgc=[11.0,12.0]$ at
redshift zero (left to right, respectively).   Color-coding differentiates the age cohorts and refers to their
formation time (see legend). }
\end{figure*}

The radial excursions seen in Figure~\ref{fig:rform} are the result of two
distinct scattering processes.  Stars can be born on, 
or non-resonantly scattered
to, eccentric orbits with large epicyclic amplitude, causing large in-plane
motions that sweep through wide ranges in radius without increasing the orbit's
angular momentum.  Stars can also can gain or lose angular momentum, thereby
changing their guiding centers, through resonant interactions \footnote{\citet{Schonrich09} refer to the latter process as ``churning'' and the former as ``blurring.''} with spiral waves
\citep[\eg,][]{Sellwood02, Roskar12a}, bar resonance overlap \citep{Minchev10},
and satellite perturbations \citep{Quillen09, Bird12}. Figure~\ref{fig:gcr} is
similar to Figure~\ref{fig:rform} but it shows the 
distribution of 
current guiding center radii
(\gcr) for each age cohort; a star with angular momentum \jz\ has a
guiding center equal to the radius of a circular orbit with the same angular
momentum. 
If a star's guiding center radius lies outside the indicated annulus,
then the star appears in the annulus only because its eccentricity
allows it to spend some of its orbit there.
For example, the \s{2.0}{4.0} population in the solar annulus
($\rgc=7.0-9.0$ kpc, yellow curve in the middle panel of 
Figure~\ref{fig:gcr}) are predominantly stars with
$R_g < 6$ kpc, which are present near the solar radius because
they are on the outer excursions of their eccentric orbits.

The trends with age in Figure~\ref{fig:gcr} reflect the combination
of inside-out formation and the age-kinematics correlation.
Older cohorts have hotter kinematics either at birth or after
early heating by mergers, so they are more likely to spend time
well outside their guiding center radius.  Furthermore, because 
older stars form preferentially at small \rgc, the members of
old cohorts found in outer annuli are primarily stars on outward
radial excursions.  By contrast, the young stars
(\s{12.8}{13.4}) in each annulus tend to be stars that were
born in that annulus and have guiding centers in that annulus.
The oldest stars (\s{0.0}{2.0}) in each annulus are generally
on highly eccentric, low angular momentum orbits with low $R_g$.

\begin{figure*}[p]
\includegraphics[width=\textwidth]{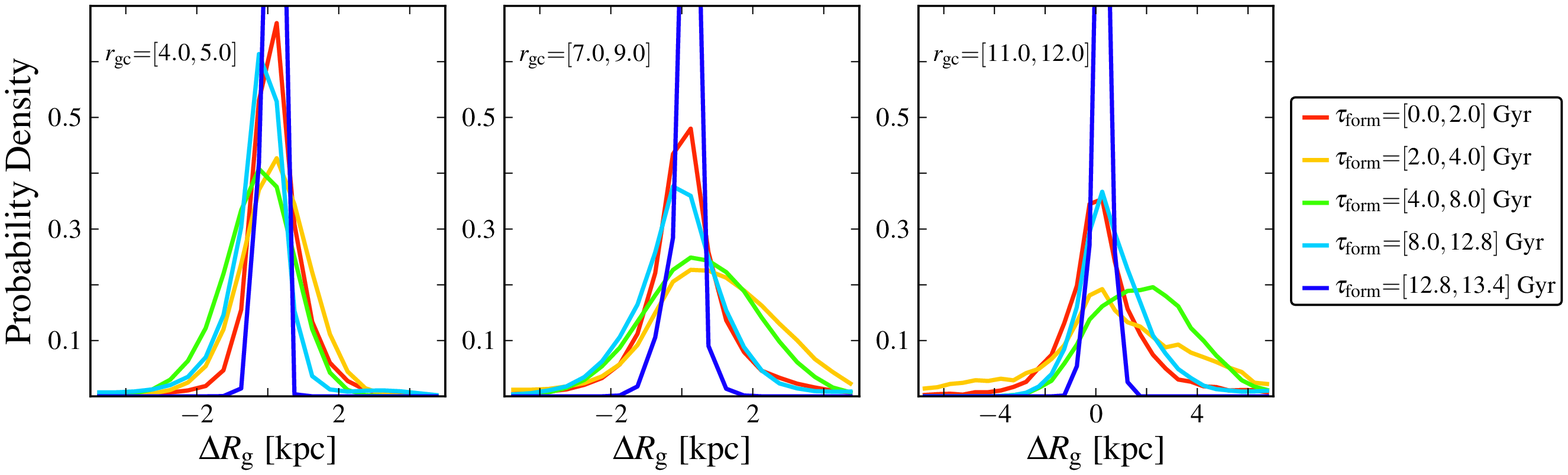}
\caption{\label{fig:dgcr}The change in guiding center radius of age cohort
members residing at $\rgc=[4.0,5.0]$, $\rgc=[7.0,9.0]$, and $\rgc=[11.0,12.0]$
at redshift zero (left to right, respectively). Color-coding differentiates the
age cohorts and refers to their formation time (see legend). Positive \dgcr\
indicates the particle has migrated outwards; particles migrating inwards have
negative \dgcr.  Note the change of horizontal scale in the right panel.}
\end{figure*}

Scattering by the corotational resonances (CR) of spiral density
waves is a particularly interesting mechanism of radial migration
because it allows stars to shift their guiding centers (change
angular momentum) without increasing orbital eccentricity
\citep{Sellwood02}.  Figure~\ref{fig:dgcr} shows the
distribution of $\Delta R_g$, the change in guiding center
radius between formation and $z=0$, for the same three annuli
illustrated in Figures~\ref{fig:rform} and~\ref{fig:gcr}.
At $\rgc=4.0-5.0$ kpc, these distributions are symmetric and
fairly compact for all age cohorts.  At the solar annulus,
the \s{2.0}{4.0} and \s{4.0}{8.0} distributions are positively
skewed and substantially broader, a trend that is still stronger
at $\rgc=11.0-12.0$ kpc.  Even the \s{8.0}{12.8} cohort has a
positively skewed $\Delta R_g$ distribution at this outer
radius.  The imposed condition of older stars at large radii
preferentially selects stars that have experienced positive $\Delta R_g$.

In sum, radial migration is an important effect in the outer galaxy.
At the solar annulus and beyond, most stars with $t_{\rm form} < 8.0$ Gyr
formed well inside their current galactocentric radius
(Figure~\ref{fig:rform}).  This migration reflects the combined
impact of the higher orbital eccentricities in older populations,
which allow stars to spend much of their orbits well beyond their
guiding center radii (Figure~\ref{fig:gcr}), and angular momentum
changes that shift guiding center radii outwards (Figure~\ref{fig:dgcr}).
The importance of radial migration in
the outskirts of disk galaxies was made clear by \citet{Roskar08a} and \citet{SanchezBlazquez09}; studies
cite radial migration as the probable origin for the inverse age-radius
gradients seen outside the break radius of nearby disk galaxies
\citep[\eg,][]{RadburnSmith12, Yoachim12}. Our findings here, in conjunction
with Figure~\ref{fig:400_den}, suggest that the older, migrating stars travel
past the break radius of the galaxy's gas reservoir at late times as traced by
the density profile of the youngest stars (\s{12.8}{13.4}). We discuss possible
implications for thick disk evolution in the Section~\ref{sec:discussion}.

\section{Discussion}
\label{sec:discussion}

The shared lifetime of mono-age populations makes them physically intuitive
descriptors of the galaxy formation process.  A cohort's members experience
relatively similar dynamical histories.  Their spatial origin
tracks the star formation process, and their initial kinematics traces that of
the gas at the cohort's formation epoch. The oldest three cohorts
(\s{0.0}{0.5}, \s{0.5}{1.0}, \s{1.0}{2.0}) describe the galaxy's turbulent
youth.  The first significant star formation occurs in the parent halo; the 
{\it in situ} component of the old cohorts show the classic signs of inside-out
galaxy formation. As gas accumulates, cools, and collapses within the main
halo, both the rate and radial extent of star formation contributing to the 
{\it in situ} population increases (Figures~\ref{fig:tf0.0_den},
~\ref{fig:tf0.5_den}, ~\ref{fig:tf1.0_den}; top left panels). The spatial
distribution of formation sites reveals a similar growth process for
satellite-born stellar populations within their host halos (same Figures as
above). During the major merger epoch, the energy in the velocity
difference of the interacting satellite and parent halos is converted into the
internal energy of the remnant \citep{BinneyTremaine}.  All cohorts that exist
at this time show dramatic and rapid increases in their internal energies (see
kinematics of \s{0.0}{0.5}, \s{0.5}{1.0}, \s{1.0}{2.0};
Figures~\ref{fig:tf0.0_rp}, ~\ref{fig:tf0.5_rp}, ~\ref{fig:tf1.0_rp}). The
present day spatial configuration and kinematic description of the
aforementioned cohorts are in place once the last of their satellite-born
members have merged with the parent halo ($t\approx2$-$4$ Gyr; see
Figures~\ref{fig:tf0.0_den} - ~\ref{fig:tf1.0_rp}). Almost $90\%$ of stars with
\tform$<2$ Gyr are associated with the spheroid in the present-day galaxy
(Table~\ref{tab:decomp}). The vast majority of stars born after $t=2$ Gyr form
in the disk (top left panel of Figures~\ref{fig:tf3.0_den} and
~\ref{fig:tf7.0_den}). Monotonic trends of position and velocity with age,
consistent with stars forming from a cooling gas reservoir, characterize the
disk's assembly.  \s{3.0}{4.0} and \s{7.0}{8.0} are illustrative examples;
progressively younger cohorts form with larger radial break radii, colder
vertical velocity distributions, and more compact vertical structure (see
initial outputs in Figures~\ref{fig:tf3.0_rp} and ~\ref{fig:tf7.0_rp}).
Younger cohorts are more susceptible to secular heating mechanisms
(Section~\ref{sec:evolution}), but any changes in population thickness and
velocity dispersion with time are not sufficient to disrupt the strong trends
between cohort kinematics and age established by each cohort's initial
properties. The galaxy retains much of its assembly history: old
cohorts are born thick (or quickly become so), while increasingly younger
cohorts form thinner structures with longer radial scales.
The galaxy forms inside-out and upside-down.

\begin{figure}
\includegraphics[width=3.7in]{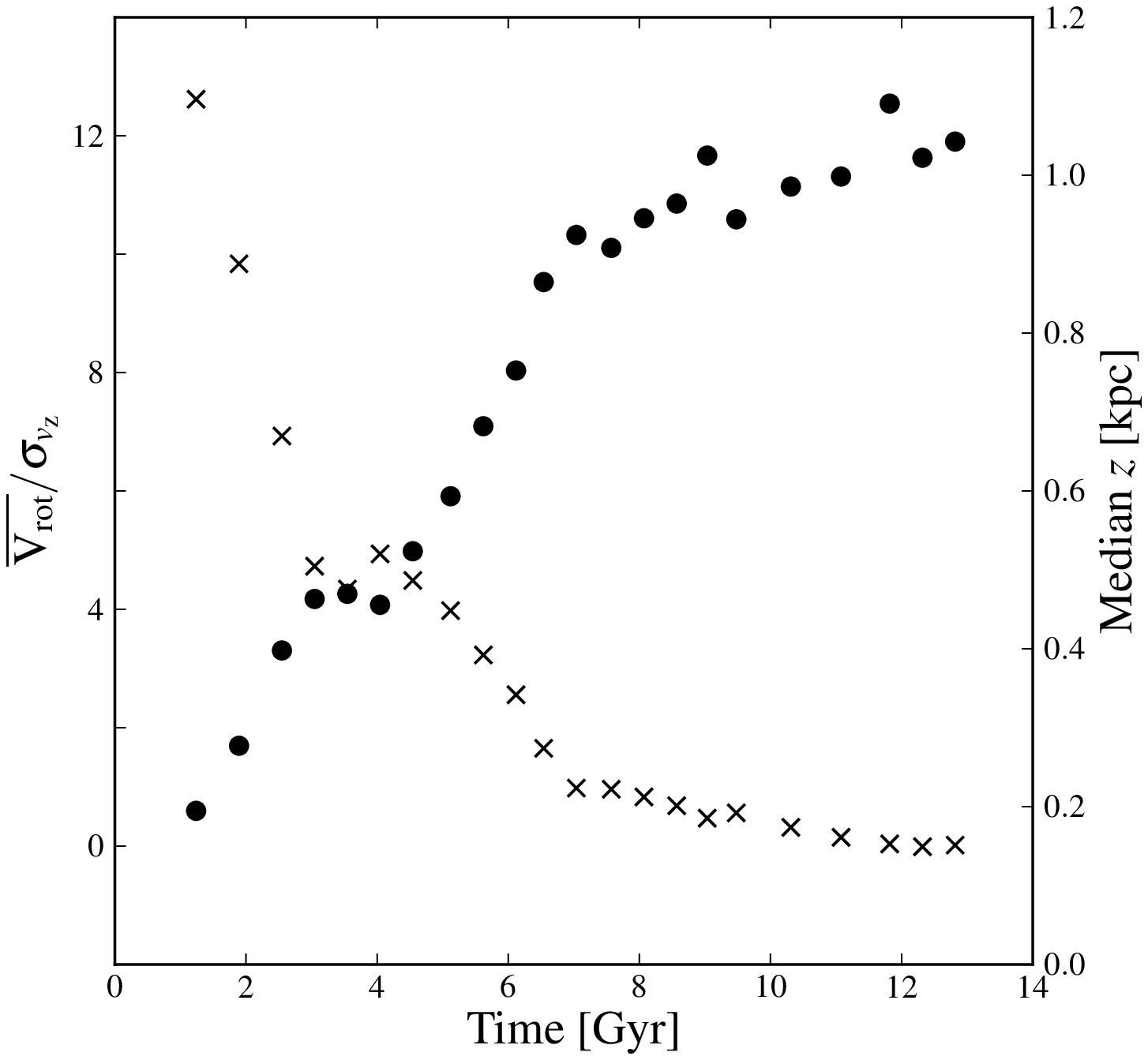}
\caption{\label{fig:vrot_vzdisp} The ratio of mean rotational velocity to vertical velocity dispersion as a function of redshift for gas under the star formation threshold temperature of the simulation ($T<3\times10^4$ K) in the disk and close to the plane ($4<\rgc<8$ kpc and $|z|<2$ kpc, filled circles). The rotated crosses mark the corresponding median height of the cold gas reservoir. Gas viable for star formation becomes increasingly rotation-dominated and vertically compact throughout the simulation. To highlight trends over noise, each marker represents a three-point time-weighted moving average.  }
\end{figure}

It is important to distinguish between a cohort's internal energy at birth and
subsequent evolution due to mergers or other scattering events. The former is
wholly dependent on the properties of the gas when the cohort forms, while the
latter is affected by gravitational interactions following the cohort's
inception. Figure~\ref{fig:vrot_vzdisp} shows the ratio of mean rotational
velocity to vertical velocity dispersion and median height for gas below the
star formation threshold temperature of the simulation ($T<3\times10^4$ K) and
spatially coincident with the growing disk ($4<\rgc<8$ kpc, $|z|<2$ kpc). Gas
eligible to form stars becomes steadily more rotationally dominated as time
progresses, leading to increasingly cooler, more vertically compact stellar
populations. \cite{Forbes12} find similar results in one-dimensional models of
the vertical structure of an evolving, star-forming disk. In our simulation,
the star-forming gas rapidly contracts despite mergers and SNe feedback at
$t<3$ Gyr due to its high physical density (and short cooling time) within the
small halo in the early universe. By $t\approx3$ Gyr, the cooling time is long
enough that strong perturbations by several mergers
(Figures~\ref{fig:tf0.5_den}, \ref{fig:tf1.0_den}, \ref{fig:tf3.0_den}) and
stirring by SN feedback (outflows are more powerful owing to higher star
formation rates that peak at $t\approx3.5$ Gyr) can prevent further  collapse
as suggested by the plateau at $t\approx3$-$5$ Gyr. After $t\approx 5$ Gyr, the
merger rate declines rapidly, and the gaseous disk undergoes another period of
collapse. The last minor merger, at $t\approx7$ Gyr, may temporarily hinder
further collapse, but the slow rate of contraction at $t>8$ Gyr is more
difficult to characterize.  At present, the cold gas has \vzdisp$\approx 15$
\kms over the radial range ($0.5<\rgc<9$ kpc).  The late-time evolution is the
most uncertain regime of Figure~\ref{fig:vrot_vzdisp}, as it is most likely
to depend on details of the SN feedback scheme and on poorly understood aspects
of ISM physics that can influence the level of turbulent motion in the
star-forming gas (see, e.g., \citealt{Pilkington11}).  The absence of molecular
cooling in the simulation could artificially suppress vertical contraction at
late times.  However, the evolution at $t < 8$ Gyr reflects the dynamic
assembly history of the galaxy and the impact of vigorous supernova feedback at
early times, so we expect the trends to remain robust to details of the ISM
modeling.  The evolving kinematics of the parent gas component indicate that
older stellar cohorts are truly born hotter than their younger counterparts,
yielding a smooth transition from thick to thin disk over time.

\begin{figure}[!t]
\includegraphics[width=3.7in]{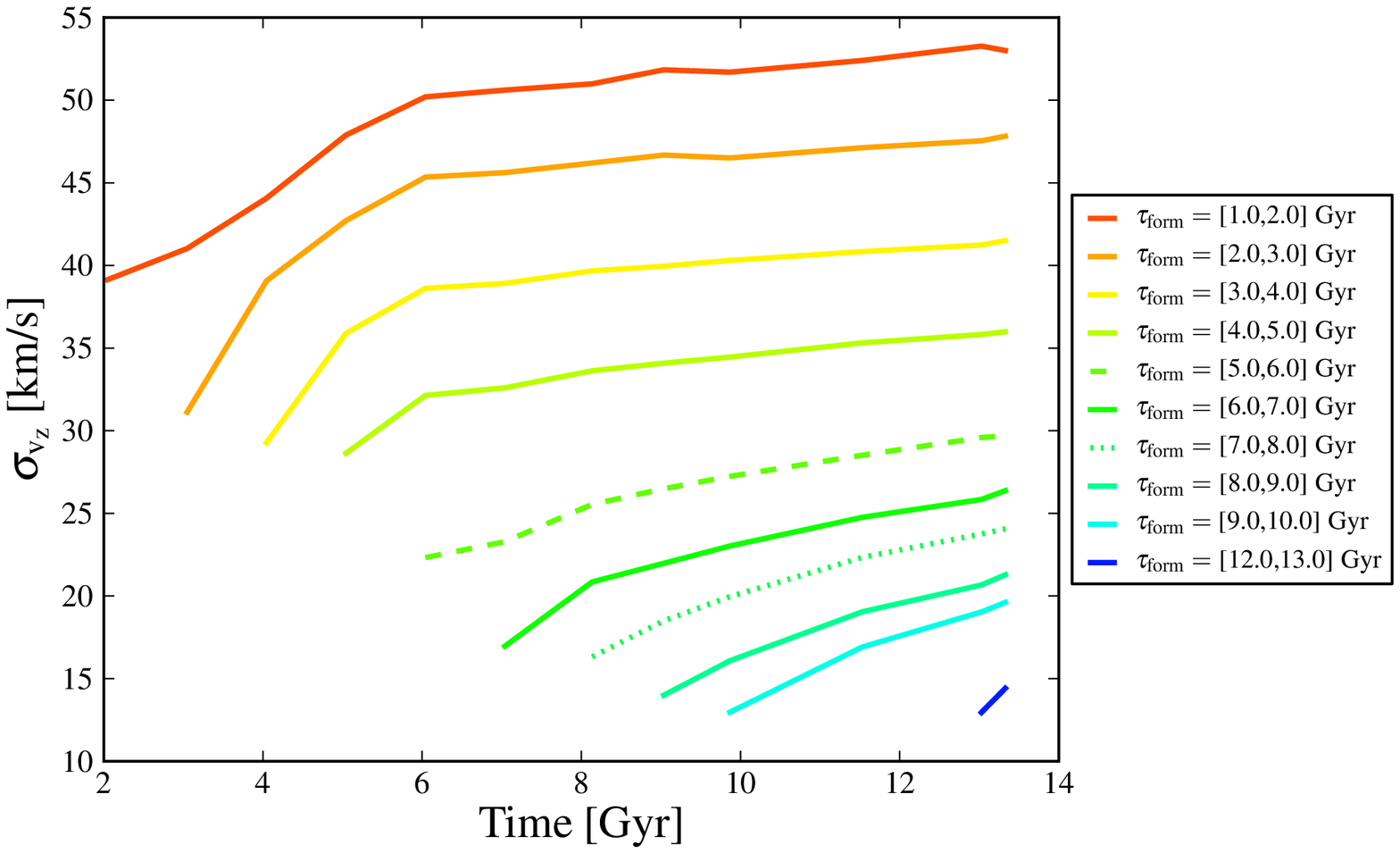}
\caption{\label{fig:temporal_vzdisp} The temporal evolution of the vertical velocity dispersion for individual cohorts. For this figure, we select likely disk orbits (\cir$>0.5$) that are spatially coincident with the disk ($4<\rgc<8$ kpc and $|z|<2$ kpc) to directly compare with Figure~\ref{fig:vrot_vzdisp}.  Color-coding
differentiates the age cohorts and refers to their formation time; line types
help distinguish amongst similar colors (see legend). }
\end{figure}

In detail, the upside-down construction of the disk and the final stellar
age-velocity relationship (AVR) are the integrated response of the age cohorts
to the contracting gas disk and dynamical heating processes. In
Figure~\ref{fig:temporal_vzdisp}, we show the temporal evolution of \vzdisp\
for a series of cohorts with 1-Gyr formation time bins. To compare with
Figure~\ref{fig:vrot_vzdisp}, we further restrict our analysis to likely disk
stars (see figure caption for selection criteria). The initial \vzdisp\ of each
cohort closely follows the properties of the star-forming gas.  The plateau
seen in Figure~\ref{fig:vrot_vzdisp} from $t\approx3$-$5$ Gyr is echoed by the
similar initial \vzdisp\ of \s{2.0}{3.0}, \s{3.0}{4.0}, and \s{4.0}{5.0}.
Similarly, initial \vzdisp\ rapidly decreases from \s{4.0}{5.0} to \s{6.0}{7.0}
as the gas reservoir contracts following the merger epoch, shows little change
during the last minor merger (compare \s{6.0}{7.0} and \s{7.0}{8.0}), and
slowly declines thereafter. For all cohorts, the heating rate
($\delta\vzdisp/\delta t$, the slope of each line) measured over any one Gyr
time interval ($t_0,t_1$) is anti-correlated with \vzdisp\ at $t_0$, confirming
that both mergers and secular scattering processes are more efficient heating
mechanisms within colder stellar populations.  Rapid increases in \vzdisp\
($>5$ \kms Gyr$^{-1}$) are temporally coincident with merger activity,
suggesting that the precise $z=0$ stellar AVR may correlate with the galaxy's
accretion history. Secular heating does affect the AVR, increasing the
dispersion of all cohorts over time, but it is clear from
Figure~\ref{fig:temporal_vzdisp} that the overall trend of \vzdisp\ with age is
dominated by differences in initial dispersion at birth rather than by
differential heating.  

There is remarkable qualitative
similarity between the age-kinematics trends measured in
the $z=0$ simulated galaxy and those observed in the MW. 
Cohort age is positively correlated with the
population's vertical extent but anti-correlated with its radial extent
(Figure~\ref{fig:400_den}).  Recent MW studies, using chemical composition as a
proxy for stellar age, suggest similar structure in our own Galaxy
\citep[\eg,][]{Bovy12a}. The alpha-enhanced, old, thick disk has a shorter
scale length than the alpha-poor, young, thin disk \citep{Cheng12b}.  Stars
far from the plane ($|z|>1$ kpc) have similar \feh\ over a broad range in
radius \citep{Cheng12a}, suggesting that these stars may share a temporal
origin. Stellar composition gradients in the vertical direction are even more
well established; observations in the solar neighborhood and beyond find that
\afe\ increases and \feh\ decreases with height above the disk plane
\citep[\eg,][]{Bensby03, Ruchti10, Lee11, Schlesinger12}. 
In addition to these geometrical differences, older cohorts in the
simulated galaxy form with higher random motions\footnote{The
trend flattens for stars with \tform$<2$ Gyr.} (Figure~\ref{fig:400_rp}).
This general age-velocity relationship (AVR) exists
in the MW as well \citep[\eg,][]{Freeman08}, and it is often explained by 
older stars having more time to be heated by various perturbing events (\eg,
molecular clouds, transient waves, satellites). 
Here we find that while the
disk-forming cohorts show a modest increase in their random motions with
time, older cohorts \emph{are born} kinematically hotter than younger stars.
Our analysis of the mono-age populations makes it clear that the galaxy's
violent youth and {\it in situ} star-formation are responsible for the
positively-sloped AVR, in addition to the age-density gradients present in
both the radial and vertical dimensions.

\begin{figure*}
\includegraphics[width=\textwidth]{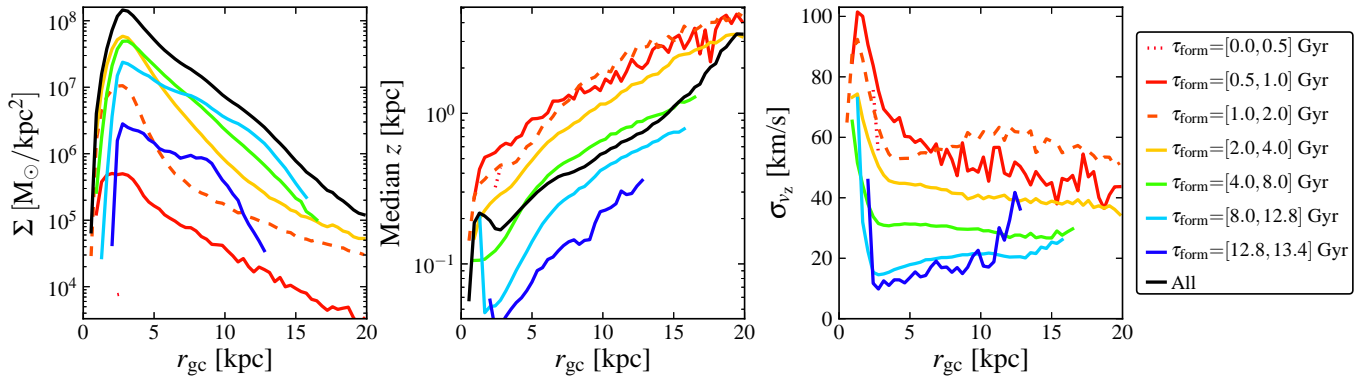}
\caption{\label{fig:circ_cut} The present-day radial profiles of surface mass density
(left panel), median height above the disk plane (middle panel), and vertical
velocity dispersion (right panel) for all age cohort members selected to have
orbits consistent with disk membership at redshift zero. Plotted members meet a
circularity cut (\cir$>0.5$)  and have total energy greater than the median
total energy of all stellar particles.  The thick, black line is
the total surface mass density (left panel) and median height (middle panel) of all selected stars. Color-coding
differentiates the age cohorts and refers to their formation time; line types
help distinguish amongst the oldest cohorts (see legend).  } 
\end{figure*}
The present-day age cohorts show near one to one correspondence with the
mono-abundance populations of \citet{Bovy12a,Bovy12c}, warranting a more direct
comparison. As SEGUE's G-dwarf sample primarily probes disk populations
\citep{Bovy12a}, we identify current disk members of the simulated galaxy and
examine their kinematic properties as a function of age
(Figure~\ref{fig:circ_cut}, see caption for selection criteria). Progressively
older cohorts have shorter scale lengths (steeper slopes in left panel of
Figure~\ref{fig:circ_cut}) and larger vertical extents (middle panel); both
monotonic trends are remarkably smooth over the G-dwarf sample's radial and
presumed age range ($6\leq\rgc\leq12$ kpc and $\tau<10$ Gyr, respectively).
\cite{Bovy12a} find that each MW mono-abundance population can be globally fit
by a single exponential in the radial and vertical directions, but in the
simulated galaxy each age cohort's vertical density profile is itself a
function of radius (Figure~\ref{fig:circ_cut}, middle panel).  The observed
mono-abundance populations show trends between vertical kinematics and age
proxy: more alpha-rich, iron-poor populations have larger vertical velocity
dispersions than their alpha-poor, iron-rich counterparts \citep{Bovy12c}. The
equivalent trend is found in the simulated galaxy, where progressively older
cohorts have larger \vzdisp\ (Figure~\ref{fig:circ_cut}, right panel).
\citet{Bovy12c} find a shallow, but negative, radial gradient in \vzdisp\ for
all mono-abundance populations. Both \s{2.0}{4.0} and \s{4.0}{8.0} show a
decrease in \vzdisp\ with radius, but the vertical velocity dispersion of stars
born in the last $5$ Gyr (\s{8.0}{12.8} and \s{12.8}{13.4}) slightly increases
with radius. These results suggest that age is a good proxy for chemical composition,  a notion independently confirmed in a recent study \citep{Stinson13}. \citet{Bovy12b,Bovy12a, Bovy12c} favor internal, secular
evolution mechanisms rather than discrete, external heating events to explain
the smooth evolution of MW kinematic properties as a function of chemical
abundance. Our simulated galaxy shows analogously smooth correlations of
stellar kinematics with age, and the trends are largely imprinted at birth or
immediately thereafter.

Thick disks are thought to be fundamental in understanding the balance of
external and internal influences in disk galaxy formation
(Section~\ref{sec:intro}). \citet{Bovy12b} showed that the mono-abundance
populations span a continuous range of scale heights at the solar neighborhood
and thereby argued that the MW does not have a distinct thick disk component.
\citet{Rix13} find that integrated vertical space density of all mono-abundance
populations masquerades as observations classically identifying the thin and
thick disk. At the solar annulus in the simulated galaxy, the vertical mass
density profiles of individual age cohorts are progressively steeper for
younger populations (Figure~\ref{fig:400_vp}, middle panel). The superposition
of all age cohorts in the solar annulus results in the familiar
double-exponential profile observed in MW star counts
\citep[\eg,][]{Gilmore83,Juric_etal08}. The two exponentially-fitted components
are not distinct in origin; for example, the \s{2.0}{4.0} and \s{4.0}{8.0}
cohorts populate both the thin and thick components at the solar annulus and
vary their fractional contributions to these structures as a function of radius
(Figure~\ref{fig:400_vp}).  Overall, our simulation leads to a ``continuous''
view of the thick disk similar to that advocated by \citet{Bovy12b}. We note
that radial migration can produce a similar vertical structure
\citep{Schonrich09b}, but in our simulation the double-exponential emerges
principally from the superposition of trends imprinted by upside-down and
inside-out formation.

The mechanisms primarily responsible for the main components of galactic
structure (determined by kinematic decomposition, Table~\ref{tab:decomp}) can
be inferred from the histories of corresponding member cohorts.  The
pseudobulge, predominantly made of stars with \tform$<4$ Gyr, formed {\it in
situ} and at early times (Figures~\ref{fig:tf0.5_den} and ~\ref{fig:tf1.0_den};
for a full investigation of bulge formation in a similar cosmological
simulation, see \citealt{Guedes12}). Over half the mass in the spheroid was
born just prior to or during the major merger epoch ($t=0.5$-$2.0$ Gyr), and
another $25\%$ formed immediately thereafter (\s{2.0}{4.0}), suggesting stellar
halo formation by stars scattered during the merger epoch, with a moderate
contribution of stars stripped from satellites.  The kinematically defined
thick disk begins to form just before the end of the major merger epoch
(\s{1.0}{2.0} accounts for $25\%$ of the thick disk's mass at $z=0$). Still,
the majority of the thick disk forms in the parent halo after the merger epoch
ends; over half the thick disk's mass is in \s{2.0}{4.0}, which has a strong
{\it in situ} component (Figure~\ref{fig:tf3.0_rp}).  MW observations may call
for a significant contribution of $t_{\rm form}=4-6$ Gyr stars in a chemically
defined thick disk \citep{Bensby04}, and in our simulation such stars are
formed overwhelmingly in the parent halo.  Even as thick disk formation is well
underway, stars with thin disk kinematics begin to appear. $26\%$ of the stars
on cold, thin disk orbits at $z=0$ are members of \s{2.0}{4.0}, but many of
these stars would likely be members of a chemically defined thick disk.  The
fraction of new stars that are kinematically associated with the thin disk
increases with time in all subsequently formed cohorts. 

Kinematic vs.\ chemical definitions of disk components likely involve
significant mixing of the thin and thick populations
\citep[\eg,][]{Navarro11,Lee11}.  However, our kinematics-based results are
broadly consistent with the chemistry-based results of \cite{Brook12}, who
study a cosmological zoom-in simulation of a disk galaxy that is $< 1/5$ the
virial mass of the MW.  They investigate the history of the gas reservoirs
eventually forming the chemically-selected thin disk, thick disk and halo to
describe galactic component formation in the annulus $\rgc=7$-8 kpc.  Their
halo stars predominantly form {\it in situ} and are scattered to the halo
during an early merger epoch, with a smaller fraction directly accreted from
satellite galaxies.  After the major merger epoch in their simulation ends,
half of the thick disk forms from hot gas that was already in the central
galaxy during the merger epoch. \citet{Brook12} report a smooth transition in
star formation from thick to thin disk stars, with the majority of their thin
disk stars forming from smoothly accreted gas after the merger epoch.
\citet{Brook12} also find that old stars must move away from their formation
radii to populate the solar vicinity.  Our analysis in \S\ref{sec:radmix_tf}
leads to a similar conclusion and differentiates between kinematically hot
orbits and guiding center changes (discussed further below).  Our simulated
galaxy and that of \citet{Brook12} form their disk components in qualitatively
similar fashions despite nearly an order-of-magnitude difference in galaxy mass
and many differences of detail in the simulations.  Both experiments have
relatively quiescent merger histories \footnote{The last major merger in
  \citet{Brook12} occurs at $z=2.7$.} by construction, so we cannot say whether
  more active merger histories would substantially change the picture and/or
  lead to disagreement with observed MW structure.

Stars radially migrate throughout our simulation, and the redistribution of
these stars impacts the galaxy's construction to an extent. For simplicity, we
focus on the role of radial migration in the creation of the thick disk at the
solar annulus, which has been a topic of debate in the recent literature
\citep[\eg,][]{Schonrich09, Sales09, Loebman11, Brook12, Minchev12a,
Roskar12b}.  The basic idea is that the hotter populations of the inner galaxy
will increase their scale heights as they move to the weaker vertical potential
of the outer disk, but there are complications because of adiabatic cooling and
dynamical selection of the stars that migrate.  We now compare the kinematics
of outwardly migrating ($\dgcr>1.0$ kpc, designated ``om'') and non-migrating
($|\dgcr|<0.5$ kpc, ``non'') disk populations \footnote{Disk stars are
  identified according to the criteria used for Figure~\ref{fig:circ_cut}.} in
  the solar annulus. When all disk stars currently residing at $7<\rgc<9$ kpc
  are considered, outwardly migrating stars and 
  non-migrators  have virtually the same vertical velocity dispersion ($\sigma_{v_{\mathrm{z,om}}}=29.7$ \kms vs.
  $\sigma_{v_{\mathrm{z,non}}}=29.5$ \kms). These results are consistent with
  those of \citet{Minchev12a}, who find that outwardly migrating stars regulate
  themselves such that they are only marginally ($<5\%$ level) hotter than the
  stellar populations at their destination radius, suggesting that migrators do
  not ``thicken'' the disk. A more complex picture emerges when we further
  dissect the sample and examine two older, thick-disk contributing cohorts
  (\s{2.0}{4.0} and \s{4.0}{6.0}). In \s{2.0}{4.0}, non-migrators have more
  vertical energy ($\sigma_{v_{\mathrm{z,non}}}=46.6$ \kms) than their
  outwardly migrating counterparts ($\sigma_{v_{\mathrm{z,om}}}=37.0$ \kms).
  \s{4.0}{6.0} shows the same relationship between migration and approximate
  vertical energy: non-migrators are hotter than outward migrators
  ($\sigma_{v_{\mathrm{z,non}}}=36.5$ \kms\ and
  $\sigma_{v_{\mathrm{z,om}}}=29.1$ \kms, respectively). 

The relationship between horizontal and vertical motions in an orbit is
well-described using vertical action ($J_z$) invariance \citep[\eg][]{Binney10,
Binney11, Schonrich12a}, and simulations show that migrating populations
conserve their vertical action on average \citep{Solway12}. Within an
idealized, one-dimensional disk galaxy, the vertical velocity dispersion of an
outwardly migrating population with constant $J_z$ scales with the surface density of the
disk, \ie, $\sigma_{v_{\mathrm{z,om}}} \propto\Sigma^{1/4}$ (assuming
$\Sigma\propto e^{- \rgc/\mathrm{r}_d}$ where r$_d$ is the scale length of
the disk; see \citet{Minchev12a} or \citet{Roskar12b} for derivation in the 1D
case, \citet{Schonrich12a} for a more general treatment). In the same model galaxy, the vertical velocity
dispersion of an isothermal, non-migrating population 
also scales with disk surface density but has an additional dependence
on the scale height of the population as a function of radius
($\mathrm{h}_z$), \ie, $\sigma_{v_{\mathrm{z,non}}}
\propto\sqrt{\Sigma\mathrm{h}_z}$ \citep[\eg,][]{Kruit11}. Comparing the radial dependence of \vzdisp\
for migrating and non-migrating populations, we find
\begin{equation} 
\frac{\vzom}{\vznon}\propto
\frac{\Sigma^{1/4}}{\sqrt{\Sigma\mathrm{h}_z}} 
\label{eq:vz}
\end{equation} 

. This 1D scaling relationship cannot be used to make quantitative predictions
of \vzom$/$\vznon, but it suggests that non-migrating stars are likely to have
more vertical energy than outward migrators at a given radius when the disk
scale length is short or h$_z$ is a strong, positively-sloped function of
radius. The simulated galaxy presents a more complicated, multi-component
scenario, but our measurements within the solar annulus disk sample (previous
paragraph) follow the relationships established in equation~\ref{eq:vz}. There,
\vznon\ and \vzom\ are nearly the same for the ensemble population, but
\vzom$/$\vznon\ decreases for \s{2.0}{4.0} and \s{4.0}{6.0}, both of which have
relatively short scale lengths and steep $\mathrm{h}_z$\footnote{The slope of
    $\mathrm{h}_z$ and the median height as a function of radius are equal if
    the if the vertical density distribution is exponential.}
    (Figure~\ref{fig:circ_cut}). 

The trend of scale height with radius may therefore play an important role in
the relative influence of migrators and non-migrators at large scale heights in
MW-like galaxies. The simulated disk sample shows increasing median height
(\zmed) with radius as $\frac{\Delta\zmed}{\Delta\rgc}=0.07$ over the radial
range $5<\rgc<14$ kpc ($71$ pc per kpc; middle panel, black line,
Figure~\ref{fig:circ_cut}).  Radial migration may contribute to this radially
increasing scale height. We find that migrating stellar populations become more
vertically extended as they move outward from the inner disk \citep[in
agreement with][]{Roskar12b}, and some old, migratory groups currently have
velocity dispersions exceeding the typical dispersion of thick disk stars in
the MW \citep[$\vzdisp=35$\kms,][]{Bensby03}.  These old stars populate both
the traditional thin and thick disk components at their destination radius, so
they are important when discussing population demographics and chemodynamic
trends.  The radial dependence of $\mathrm{h}_z$ observed here is contrary to
the traditional notion of constant scale height in disk galaxies
\citep[\eg,][]{Kruit82}, but radially increasing stellar scale heights have
been found in more recent measurements of a large sample of nearby disk
galaxies \citep[][; $\frac{\Delta\mathrm{h}_z}{\Delta\rgc}=0.05$ for MW Hubble
type]{deGrijs97} and in the MW itself \citep[][;
$\frac{\Delta\mathrm{h}_z}{\Delta\rgc}=0.02$]{Narayan02}.  Simulation
improvements, including incorporating missing ISM physics such as molecular
cooling, will likely flatten $\mathrm{h}_z$ and decrease the tension between
simulations and observations.  Individual age cohorts may still show steep
$\mathrm{h}_z$ within the disk because of heating events like mergers, as even
minor perturbations can cause flaring due to the low restorative force in the
outer disk \citep[\eg,][]{Kazantzidis08,Bournaud09}.  Old, thick disk
contributing cohorts (\eg, \s{2.0}{6.0}) in MW-sized halos are particularly
susceptible to flaring: these stars are typically born during a significant
merger period (Section~\ref{sec:evolution}) and non-migrators at the solar
annulus would form at the edge of a young, relatively compact disk . If this
scenario is generic to older age cohorts in $M_\star$ halos, the maximal
vertical extent of the thick disk (and similar structures in external galaxies)
is likely set by the initially hot, {\it in situ} stars, as measured in the
simulated galaxy.

\section{Conclusions}
\label{sec:summary_tf}

We have investigated the structure, kinematics, and evolutionary history of
stellar age cohorts in one of the highest resolution cosmological simulations
ever run of the formation of a MW-like galaxy.  This simulation produces good
(though not perfect) agreement with many observed properties of the MW and
similar galaxies \citep{Guedes11}, so it may provide useful guidance to the
origin of trends found in the MW.  We find remarkably good qualitative
agreement between the trends for our mono-age cohorts and the observed trends
for mono-abundance stellar populations in the MW found by \citet{Bovy12b,
Bovy12a, Bovy12c}.  Within the disk, older cohorts have shorter radial scale
lengths, larger vertical scale heights, and hotter kinematics.  The trends with
age are smooth, but at a given radius the superposition of old,
vertically-extended populations and young, vertically-compact populations gives
rise to a total vertical profile that has the double-exponential form usually
taken to represent the superposition of ``thin'' and ``thick'' disks in the MW.

Tracing back the formation history of each age cohort, we find that these
present-day trends of spatial structure and kinematics are largely imprinted at
birth.  The oldest stars (\s{0.0}{0.5}, \s{0.5}{1.0}, and, to a lesser extent,
\s{1.0}{2.0}) form during the active merging phase of the galaxy's assembly,
and these mergers scatter stars into rounded configurations and kinematically
hot orbits, with the oldest stars being the most centrally concentrated.  Even
in these early cohorts, the great majority of stars form in the main halo
rather than satellites, and phase-mixing erases differences in their
distributions in any case.  Younger cohorts form in disk-like configurations
that are progressively more extended, thinner, and colder as $t_{\rm form}$
increases.  The larger scale heights at earlier times are mostly inherited from
the star-forming gas disk, which has a greater degree of turbulent support
relative to rotational support (Figure~\ref{fig:vrot_vzdisp}), presumably
because of stronger SN feedback from more vigorous star formation and more
dynamical stirring by mergers and halo substructure.  A disk cohort typically
shows additional heating for another $1-2$ Gyr after $t_{\rm form}$, but
thereafter the growth of velocity dispersion or vertical thickness is minimal.

The ``inside-out'' radial growth of the disk is a long-standing theoretical
expectation --- even if the gas disk were fully present at $t=0$,
star-formation would proceed more rapidly in the higher surface density,
central regions.  The resulting shape of the radial abundance gradient early in a galaxy's evolution can therefore directly constrain feedback models \citep{Pilkington12,Gibson13}. At redshift zero, inside-out growth readily explains the direction of radial
abundance gradients measured both locally and in external disk galaxies and is
a central component of many Galactic chemical evolution models
\citep[\eg][]{Chiappini97, Prantzos00}. Radial migration of stars is also an
important ingredient in understanding chemical evolution
\citep{Wielen96,Sellwood02,Roskar08b,Schonrich09}, and it arises naturally in
our simulation.  In the outer disk, progressively older stars have
progressively smaller formation radii on average, though the distribution of
$r_{\rm form}$ is usually broad and includes some old stars born at large
radius.  The appearance of old stars at $r_{\rm gc} > r_{\rm form}$ is partly a
consequence of increases in angular momentum (and thus guiding center radius)
and partly a consequence of the larger orbital eccentricities of older stellar
populations.

The ``upside-down'' evolution that we find for the disk's vertical structure in
our simulation is more novel, and it highlights a view of the MW's thick disk
that has growing support from analytic and numerical galaxy formation studies.
In our simulation, the thick disk arises from continuous trends between stellar
age and the evolving structure of the star-forming disk, not from a discrete
merger event or from secular heating or radial migration after formation.  The
simulation clearly reproduces the basic phenomenology of the MW's thin and
thick disks: a double-exponential vertical profile, and an increasing fraction
of old stars at large distance from the plane and/or large random velocity.
\citet{Forbes12} have proposed a similar model for the origin of the thin/thick
transition, based on one-dimensional simulations that track the evolution of
turbulence in the star-forming gas. ``Upside-down'' formation is also observed
to varying degrees in gas-rich merger simulations \citep{Brook04}, zoom-in
cosmological simulations \citep{Stinson13}, and in clumpy, unstable gas-rich
disks \citep{Bournaud09}. The disk heating study of \citet{House11} found, much
like our study, that upside-down formation can naturally lead to a smooth
stellar AVR at redshift zero. Quantitative tests of the simulation predictions
will require using chemistry in place of age and carefully matching the
selection criteria and measurement uncertainties of observational samples, a
task that we reserve to future work.

Any simulation of galaxy formation has numerical limitations and physical
uncertainties associated with the treatment of star formation and feedback.
The specifics of our star formation treatment may matter for some of our
conclusions, as we find that the thickness of the stellar disk is largely
inherited from that of the star-forming gas.  The Eris simulations have many
successes in reproducing disk galaxy properties at high redshift \citep{Shen12}
and low redshift \citep{Guedes11}, but the adopted star formation model may
predict galaxies that are too luminous for their halo mass relative to
abundance matching constraints \citep{Moster13, Munshi13}.  Solving this
problem may require simulations with more vigorous feedback, and it will be
important to confirm that the trends found here persist in such simulations.
Furthermore, this simulation is a single realization selected to have a
quiescent merger history, which limits our ability to draw statistical
conclusions.  However, we find good qualitative agreement with the simulation
of \cite{Brook12} despite a roughly order-of-magnitude difference in galaxy
mass,  which suggests that our conclusions are fairly robust, and upside-down
formation also arises in the one-dimensional calculations of \cite{Forbes12} as well as a small but diverse set of numerical experiments \citep[\eg][]{Brook04, Bournaud09, House11, Stinson13}. The good agreement between the properties of mono-age populations and observed trends in the MW further suggests that this
recipe for galaxy formation has most of the right ingredients.

ELS envisioned a monolithic collapse of the proto-Galaxy, a scenario at odds
with the hierarchical formation predicted in the CDM model.  Our simulated
galaxy forms self-consistently via violent, hierarchical accretion and cold
flows at early times \citep{Shen12} followed by a nearly quiescent phase in
which the disk grows predominantly from cold flow accretion and, to a lesser
extent, from cooling of a diffuse warm/hot corona \citep{Brooks09, Guedes11}.
Nonetheless, there are some qualitative similarities between our simulation
results and the ELS scenario: the bulge \citep{Guedes12} and the thin and thick
disks (this paper) form the overwhelming majority of their stars {\it in situ},
and these stars form in progressively more flattened, more rotationally
supported configurations.  The global distributions and correlations of galaxy
properties --- luminosities, colors, rotation speeds, morphologies, etc. ---
provide critical tests of galaxy formation simulations in the $\Lambda$CDM
cosmology.  As the realism and resolution of simulations continues to improve,
and as chemodynamical surveys of the MW grow in scope and detail, the
close-to-home tests of simulation predictions against the observed correlations
of the age, composition, spatial structure, and kinematics of stellar
populations will provide steadily more powerful constraints on the theory of
galaxy formation.

\acknowledgements We thank Ralph Sch\"onrich and the referee, Brad Gibson, for
detailed comments on the original manuscript that led to significant
improvements in the paper. We acknowledge fruitful conversations with Tom
Quinn. We made use of pynbody (http://code.google.com/p/pynbody) in our
analysis for this paper. J.B. acknowledges the support of the Vanderbilt Office
of the Provost through the Vanderbilt Initiative in Data-intensive Astrophysics
(VIDA) and the Distinguished University Fellowship awarded by the Ohio State
University. S.K. is supported by the Center for Cosmology and Astro-Particle
Physics at The Ohio State University. This research was partially funded by the
Marie Curie Actions for People COFUND Program through the ETH Zurich
Postdoctoral Fellowship awarded to J.G. Support for this work was provided by
the NSF through grants AST-0908910, AST-1009505, AST-1211853, and OIA-1124453,
and by NASA through grant NNX12AF87G. We acknowledge the Ohio Supercomputer
Center (http://www.osc.edu) for computing time.

 
\newcommand{\noopsort}[1]{}

\end{document}